\documentclass[]{aa}
\usepackage{graphics,times}
\begin{document}
%\thesaurus{11(03.13.3;  % Methods, miscellaneous 
%           03.13.4;  % Methods, numerical
%           11.11.1;  % Galaxy: kinematics and dynamics
%           11.19.2;  % Galaxies: spiral
%           )}
\title{Discrete Dynamical Classes For Galaxy Discs\\ 
and the implication of a second generation of \\ Tully-Fisher Methods}
\author{D. F. Roscoe}
\offprints{D. F. Roscoe}
\institute{School of Mathematics, \\
Sheffield University, Sheffield, S3 7RH, UK. \\
Email: D.Roscoe@shef.ac.uk \\
Tel: 0114-2223791, Fax: 0114-2223739}
\date{Received 1 June 1999 / Accepted 1 September 1999}
\abstract{In Roscoe \cite{RoscoeA}, it was described how the modelling of a small 
sample of 
optical rotation curves (ORCs) given by Rubin et al \cite{Rubin} with
the power-law $V_{rot}=AR^\alpha$, where where the parameters $(A,\alpha)$ vary 
between galaxies, raised the hypothesis that the parameter $A$ (considered in 
the form $\ln A$) had a preference for certain discrete values.
This specific hypothesis was tested in that paper against a sample of 900 spiral
galaxy rotation curves measured by Mathewson et al \cite{Mathewson1992}, 
but folded by Persic \& Salucci \cite{Persic1995}, and was confirmed on this large 
sample with a conservatively estimated upper bound probability of $10^{-7}$ 
against it being a chance effect.
In this paper, we begin by reviewing the earlier work, and then describe the
analyses of three additional samples; 
the first of these, of 1200+ Southern sky ORCs, was published by Mathewson \& 
Ford \cite{Mathewson1996}, the second, of 497 Northern sky ORCs, is a composite sample
provided by kind permission of Giovanelli \& Haynes published in the sequence of
papers Dale et al \cite{Dale1997}, \cite{Dale1998}, \cite{Dale1999} and 
Dale \& Uson 
\cite{Dale2000},
whilst the third, of 305 Northern sky ORCs, was published by 
Courteau \cite{Courteau}.
These analyses provide overwhelmingly compelling confirmation
of what was already a powerful result.
Apart from other considerations, the results lead directly to what can be
described as a {\lq second generation of Tully-Fisher methods.'}
We give a brief discussion of the further implications of the result.
\keywords{
rotation curves -- spiral galaxies -- galactic evolution -- discrete classes --
disc formation -- disk formation -- galaxy discs -- galaxy disks - 
Tully-Fisher}}
\titlerunning{Discrete Dynamical Classes For Galaxy Discs}
\maketitle
\section{Introduction}
\label{Intro.sec}
This paper describes the analyses of four large optical rotation curve samples to show how 
the hypothesis that {\lq spiral galaxies are constrained to occupy discrete 
dynamical classes'} is supported by the data as a statistical certainty. 
At a practical level, the result has considerable immediate implications for 
the absolute
determination of zero points for classical Tully-Fisher methods and, in the wider
context of the overall analysis, provides a class of second generation
Tully-Fisher methods in which absolute Tully-Fisher calibrations {\it and}
linewidth determinations can, in principle, be simultaneously determined on any
given sample of ORCs.
At a deeper level, the basic result appears to
have implications for our understanding of galaxy evolution.
The result is so unexpected, that a short review of already published
material (Roscoe \cite{RoscoeA}) is likely to be useful to the present reader.

Arguments based on certain symmetry considerations - the nature of which are 
not immediately relevant here - led us to consider the possibility that the 
segments of optical  
rotation curves which occupy disc regions (given an operational definition 
later in this text)
might be reasonably described - in an overall statistical sense - by power laws in the 
form $V_{rot} \,=\,A R^\alpha$, with the parameters $A$ and $\alpha$ being determined empirically
for each galaxy in turn.
As a means of gaining familiarity with this idea, we considered
the small sample of 21 ORCs published by Rubin et al \cite{Rubin}
from this point of view.
Of this sample of 21 ORCs, only twelve manifested reasonably monotonic behaviour
and so were selected {\it on these grounds alone} as reasonable candidates
for a power law analysis.
Subsequently, 
a linear regression of the model $\ln V_{rot} \,=\, \ln A \,+\,\alpha \ln R$
onto each of the twelve ORCs provided twelve sets of 
parameter-pairs $(\alpha,\,\ln A)$.
The first clear result of this mini-analysis was
that $\alpha$ and $\ln A$ appeared to be very strongly correlated - and this 
particular aspect has now been analysed in detail using Persic 
\& Salucci's \cite{Persic1995} folding solution for 900 ORCs from the 
Mathewson et al \cite{Mathewson1992} sample (Roscoe \cite{RoscoeB}).

However, as reference to table \ref{Table1} shows (the entries of which have 
been rounded to the nearest decimal), a curious numerical 
coincidence arose - specifically, that
every one of the twelve $\ln A$ values lay between $\pm 0.15$ of an integer
or half-integer value - a coincidence that has odds around 1:500 of being a
chance occurrence ( a-posteriori probabilities\,!). 
Of course, the integer/half-integer values themselves can be of no possible
significance since, if Rubin et al \cite{Rubin} had estimated distance 
scales using a value of $H$ 
significantly different from the $50$km/sec/Mpc they actually used, then a 
completely different set of $\ln A$ values would have resulted.
So, the coincidence was simply that of regularity in spacing which would
probably have not been noticed with, say, $H\,=\,70$km/sec/Mpc.
Anyway, curiosity provided a sufficient motivation to consider the matter
further, using the Persic \& Salucci \cite{Persic1995} sample of 900 folded 
ORCs.
\begin{table}
\begin{minipage}{4in}
\caption{Twelve Rubin, Ford \& Thonnard \cite{Rubin} spirals}
\label{Table1}
\begin{tabular}{lclc}
\hline
Galaxy & $\ln A$ & Galaxy & $\ln A$ \\
\hline
N3672 & 3.6  & U3691 & 3.6 \\
N3495 & 4.0  & N4605 & 4.0 \\
I0467 & 4.1  & N0701 & 4.1 \\
N1035 & 4.1  & N4062 & 4.5 \\
N2742 & 4.5  & N4682 & 4.5 \\
N7541 & 4.6  & N4321 & 4.9 \\
\hline
\end{tabular}
\end{minipage}
\end{table}
\begin{table}
\begin{minipage}{4in}
\caption{$\ln A$ data}
\label{Table2}
\begin{tabular}{ccc}
\hline
RFT & Pred value&  Actual value\\
scale  &with MFB&MFB  \\
    & scale & scale \\
\hline 
3.5 & 3.81 & 3.85 \\
4.0 & 4.22 & 4.24 \\
4.5 & 4.63 & 4.72\\
5.0 & 5.04 & 5.06 \\
\hline
\end{tabular}
\end{minipage}
\end{table}
This sample had its distance-scaling determined by a 
Tully-Fisher
relationship calibrated by Mathewson et al \cite{Mathewson1992} against Fornax
using $H \,=\,85$km/sec/Mpc, so that the 
integer/half-integer hypothesis for $\ln A$ is not appropriate.
However, a simple analysis (described in Appendix B of Roscoe \cite{RoscoeA}, 
and 
relying on the investigation of the $(\alpha,\,\ln A)$ correlation
given in Roscoe \cite{RoscoeB}) reveals the relation
\begin{displaymath}
\ln A_{MFB} \, \approx \, 0.82\, \ln A_{RFT} \,+\,0.94
\end{displaymath}
where $A_{MFB}$ denotes the value of $A$ determined using the 
Mathewson et al \cite{Mathewson1992} scaling,
whilst $A_{RFT}$ denotes its value determined using the Rubin et al \cite{Rubin} 
scaling.
Using this latter relation, the integer/half-integer values of $\ln A$ in
the Rubin et al \cite{Rubin} scaling transform into their corresponding 
value in the Mathewson et al \cite{Mathewson1992} scaling
according to the first two columns of table \ref{Table2}.
\begin{figure}[htp]
\centering
\resizebox{\hsize}{!}{\includegraphics{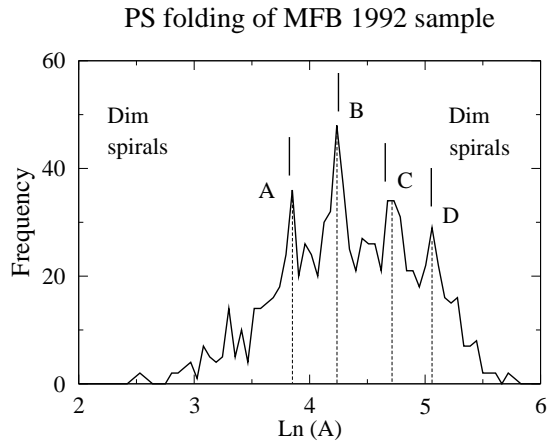}}
\caption{$\ln A$ distribution for the Mathewson et al \cite{Mathewson1992}
sample with Persic \& Salucci folding.
Vertical dotted lines indicate actual peak centres. Vertical solid 
bars indicate {\it predicted} peak centres from the Rubin et al \cite{Rubin}
sample. Bin width = 0.055.}
\label{fig1A}
\end{figure}
The actual $\ln A$ distribution of the 900 ORCs of the 
Mathewson et al \cite{Mathewson1992}
sample folded
by Persic \& Salucci \cite{Persic1995} is given in figure \ref{fig1A}.
The vertical solid bars indicate the {\it predicted} positions of the peaks,
based on our analysis of the Rubin data,
as in the second column of table \ref{Table2}, whilst the vertical dotted
lines indicate actual peak centres, as in the third column of table
\ref{Table2}.
The correspondence between the peak positions, predicted on the basis
of the twelve Rubin et al \cite{Rubin} galaxies of table \ref{Table1}, 
and the actual peak
positions is clearly remarkable.
A crude, but extremely conservative, upper bound estimate 
of the probability of the peaks in 
the distribution of figure \ref{fig1A} occurring by chance, given the original 
hypothesis defined on the small Rubin et al \cite{Rubin} sample, was
given in Roscoe \cite{RoscoeA} as $10^{-7}$.

The implications of this result are so potentially significant, that it has 
become essential to test the specific hypothesis against new samples.
This is done using three additional samples in the following sections, and the
results are overwhelmingly in favour of the hypothesis.
That is, it very much appears as though we are seeing evidence for discrete
dynamical classes in spiral galaxy discs.
\subsection{Organization of paper}
The details of the samples used are described in \S\ref{Sec2}, whilst essential computational
details are described in \S\ref{Sec4}. 
Sections \S\ref{MFB.sec}, \S\ref{MF.sec}, \S\ref{Sec.DGHU} and 
\S\ref{Courteau.sec} describe the
core analyses of the four samples, whilst the statistical analysis
of the results of these analyses is described in \S\ref{Odds.sec}.
We discuss the implications of the analysis for Tully-Fisher methods in
\S\ref{TF}. 
Major stability issues are discussed in \S\ref{Stability} whilst
a brief discussion of possible theoretical implications
is given in \S\ref{GenDynImp}.
The whole is summarized in \S\ref{Conclusions.sec}.
Various secondary issues are covered in the appendices, and referred to as necessary.
\section{The Samples}
\label{Sec2}
The basic relevant characteristics of the four samples analysed are given in 
Table \ref{Table3}, listed in order of probable quality as judged by either
mean apparent magnitude, or by \% of late-type spirals (which have a higher
hydrogen content than early-type spirals, and are therefore likely to be
associated with more accurate $H_\alpha$ Doppler shift measurements).
We discuss, and analyse, the samples in order of likely quality.

There is an additional potential problem pointed out by Bosma \cite{Bosma} in the 
particular context of the sample of Mathewson et al \cite{Mathewson1992}.
Specifically, he points out that about 17\% of the Mathewson et al
galaxies are seen edge-on and that, for these galaxies, internal
absorption effects mean that ORC measurements are subject to relatively large
errors. Table \ref{Table3} lists the \% of edge-on galaxies in each of the
samples. We discuss the effects of excluding these edge on galaxies when the
detailed analyses of the samples are given.
\subsection{The Original Sample, Mathewson et al \cite{Mathewson1992}}
\label{Sec2.1}
In the period 1988-90, Mathewson et al \cite{Mathewson1992}
measured $H_\alpha$ 
and $N_{II}$ rotation curves for 
965 Southern sky spirals on the $2.3m$ telescope at Siding Spring Observatory,
whilst the corresponding $I$-band photometry was obtained using the $1m$ and 
$3.9m$ telescopes.
The $N_{II}$ observations were used to provide an estimate of the internal
measurement accuracy of the $H_\alpha$ observations, and these estimates,
in the form of a parameter varying on the range $(0,\,1)$, were provided for 
each velocity measurement on every ORC.
So far as we are aware, the $N_{II}$ observations have not been made available.

Persic \& Salucci took this sample of 965 ORCs and subsequently produced a 
sample of 900 good-to-excellent quality folded ORCs (Persic \& Salucci
\cite{Persic1995}), suitable for 
their purpose of modelling the internal dynamics of spiral galaxies.
It was on this sample that the {\lq discrete dynamical classes'} hypothesis was 
originally tested (Roscoe \cite{RoscoeA}), and which is represented in figure
\ref{fig1A}. 
We note that about 17\% of the sample consists of edge-on galaxies.
\begin{table*}
\caption{Comparison of the four samples}
\label{Table3}
\begin{tabular}{lllccll}
\hline
       &        & Mean    &      Mean  & \% Late & \% Edge     &           \\
Sample & Sample &distance &   apparent & type    &on &  Telescope \\
       & size   & km/sec  &   magnitude& spirals & discs     &diameter \\
\hline
MFB 1992 &900   &  3651   &12.3 ($I$-band)& 43\% &17\% &  $2.3m$ \\
DGHU 1997&497   &  13747  &15.0 ($I$-band)&57\%  &11\% &$4m$, $5m$\\
MF 1996  &1200  &  6311   &13.4 ($I$-band)&18\%  & 6\% &$2.3m$ \\
SC 1997  & 305  & 5854    &13.5 ($R$-band)&45\%  & $<$\,1\%& $2.5m$, $3m$ \\
\hline
\end{tabular}
\end{table*}
\subsection{The Second Sample, Dale, Uson, Giovanelli \& Haynes \cite{Dale1997} et seq}
The second sample of 497 ORCs was provided by the kind permission
Riccardo Giovanelli and Martha Haynes specifically for the present analysis,
and is a composite of samples published by Dale, Giovanelli \& Haynes 
\cite{Dale1997}, \cite{Dale1998}, \cite{Dale1999} and 
Dale \& Uson \cite{Dale2000}, and 
was originally selected for studies of peculiar velocity distributions. 
As occasion demands we shall refer to the sample as either that of Dale et al,
or that of DGHU.
The rotation curve measurements were done on the 
Mt Palomar $5m$ telescope and the CTIO $4m$ telescope, whilst the $I$-band
photometry was done on the KPNO and CTIO $0.9m$ telescopes.

Although this sample is, typically, three to four times more distant that the 
objects of the first sample the ORC determinations were made with much larger 
telescopes,
and the sample has a very high proportion of late-type objects.
We can therefore expect that the quality of the sample will bear comparison with
both of the Mathewson et al samples.
We note that about 11\% of this sample consists of edge on galaxies.
\subsection{The Third Sample, Mathewson \& Ford \cite{Mathewson1996}}
The third sample of 1200+ ORCs was obtained by Mathewson \& Ford 
\cite{Mathewson1996} in the period
1991-93 as part of the same observing programme that gave the original
965 ORCs of Mathewson et al \cite{Mathewson1992}.
The main differences between the Mathewson \& Ford \cite{Mathewson1996} and 
Mathewson et al \cite{Mathewson1992} samples are given in table 
\ref{Table3}:
It is clear that the Mathewson \& Ford \cite{Mathewson1996} sample is, on average, $73\%$ 
more distant than 
the Mathewson et al \cite{Mathewson1992} sample, meaning that, on average, we 
only receive 1/3 as much 
light (all other things being equal) from each of the objects; this is 
consistent with the fact that there is an average of a 1.1 apparent magnitude 
difference between the samples.
This large difference in {\lq light received'} indicates that we can expect
ORC measurements on the Mathewson \& Ford \cite{Mathewson1996} sample to be significantly 
less accurate than 
those on the Mathewson et al \cite{Mathewson1992}.

Furthermore, the Mathewson \& Ford \cite{Mathewson1996} sample consists of 
only 18\% late-type spirals compared to 43\%, 57\% and 45\% respectively for the
other three samples.
Since 
late-type spirals are significantly richer in hydrogen than are early-type 
spirals, and since ORCs are measured primarily in $H_\alpha$, then we can
expect the quality of velocity measurements in the Mathewson \& Ford 
\cite{Mathewson1996} sample to rank behind that of the other samples for
this reason also.
We note that only about 6\% of this sample consists of edge on galaxies.
\subsection{The Fourth Sample, Courteau \cite{Courteau}}
The fourth sample, of 305 ORCs, was selected by Courteau from a sample
of $Sb,\,Sc$ field galaxy ORCs (Courteau \cite{Courteau}) for a linewidth/Tully-Fisher
study, and differs from the first three samples in being the only sample
with $R$-band photometry.
% Must see the SC 1997 paper for discussion of the R-band photometry
The original observations were made using the Shane $3m$ telescope at Lick
Observatories and the du Pont $2.5m$ telescope at Las Palmas.

As reference to table \ref{Table3} shows, the Courteau \cite{Courteau} sample is almost 
as distant as is the Mathewson \& Ford \cite{Mathewson1996} sample, but it contains a similar 
proportion of late-type spirals to that contained in 
Mathewson et al \cite{Mathewson1992}.
Also, the telescopes used were a little larger and so, all other things being
equal, 
we would expect the quality of these ORCs to be midway between that
of Mathewson et al \cite{Mathewson1992} and Mathewson \& Ford 
\cite{Mathewson1996} and 
similar overall to the Dale et al sample.
However, the sample is considerably smaller than the others so that statistics
performed upon it will be correspondingly less significant.
We note that less than 1\% of this sample consists of edge on galaxies.
\section{Miscelleneous computational details}
\label{Sec4}
\subsection{Pre-folding data filter}
\label{PFDF}
Persic \& Salucci \cite{Persic1995}
found that an ORC could only be folded with sufficient accuracy for their purpose if
individual velocity measurements not satisfying a pre-determined accuracy 
condition were rejected.
This is also our experience with the auto-folder method described in Roscoe \cite{RoscoeC}.
Accordingly, an individual velocity measurement on any given ORC is retained 
only if it has an estimated absolute error $\leq 5\%$.
For the Mathewson et al \cite{Mathewson1992}, Mathewson \& Ford \cite{Mathewson1996},
Dale et al \cite{Dale1997} et seq and Courteau \cite{Courteau} samples
this process leads to losses of $35\%,\,25\%,\,46\%$ and $46\%$ respectively of all
individual velocity measurements.
The folding process for each ORC is performed once this data-filtering process
is completed.
But the data-filtering itself inevitably means that some ORCs are left with
insufficient velocity points to permit subsequent accurate folding.
The attrition rates for ORCs lost to the overall analysis via this process 
are given by $3\%,\,4\%,\,8\%$ and $7\%$ for each of the four samples 
respectively.
\subsection{The computation of $\ln A$}
The {\lq discrete dynamical classes'} hypothesis is a statement which 
{\it specifically} concerns the values assumed by the set of $\ln A$ parameters,
computed for each folded ORC in turn.
It is therefore necessary to state clearly how this parameter is computed.
The following discussion assumes that each ORC is folded and corrected for 
inclination as a matter of course.

In the original mini-analysis of the 21 ORCs of Rubin et al (1980), we rejected
nine on the grounds that they were strongly non-monotonic, and therefore not 
amenable to a power-law analysis. However, the application of any such 
subjective culling procedure to the present large-scale analysis would seriously 
compromise its objectivity, and so some algorithmically defined {\lq black-box'} 
technique is required to perform the equivalent task.

Rather than trying to identify complete ORCs which are unsuitable in some way 
for our 
analysis, we note that most non-monotonic behaviour is on the interior parts 
of ORCs and, accordingly, develop an algorithmic statistical technique to 
identify {\lq bad'} 
interior sections, say $0 \, \leq\,R\,<\,R_{min}$, where they exist. 
The whole of these {\lq bad'}
regions are then rejected and the whole remaining ORC is used in
our analysis without further processing.
Apart from the ORC attrition arising from the pre-folding data-filter,
the four ORC samples are used in their entirety.
\subsection{The algorithm}
\label{TheAlgorithm}
The process to be described uses the techniques of linear regression and,
following conventional definitions, an observation is reckoned
to be {\it unusual} if the predictor is unusual, or if the response is unusual.
For a $p$-parameter model, a predictor is commonly defined to be unusual if its 
leverage $ > 3 p/N$, when there are $N$ observations. In the present case,
we have a two-parameter model so that $p = 2$.
Similarly, the {\it response} is commonly defined to be unusual if its 
standardized residual $> 2$.
The computation of $(\alpha, \ln A)$, for any given folded and
inclination-corrected rotation curve, can now be described as follows:
\begin{enumerate} 
\item Form an estimate of the parameter-pair $(\alpha, \ln A)$ by 
regressing the $\ln V_{rot}$ data on the $\ln R$ data for the folded ORC; 
\item Determine if the {\it innermost} observation only is an {\it unusual} 
observation in the sense defined above;
\item If the innermost observation is unusual, then exclude it from the
computation and repeat the process from (1) above on the reduced data-set;
\item If the innermost observation is {\it not} unusual, then no further 
computation is required - the current values of $(\alpha,\ln A)$ are considered
as final.
\end{enumerate}
This algorithm has the result that, on average, $(\alpha,\ln A)$ is computed on
the exterior 88\% of the data points in each ORC of the Mathewson et al 
\cite{Mathewson1992}
sample, the exterior 87\% of data points in each ORC of the Mathewson \& Ford 
\cite{Mathewson1996} sample, the exterior 91\% of data points in each ORC of the
Dale et al \cite{Dale1997} et seq sample and the exterior 91\% of the data 
points in each ORC of the Courteau \cite{Courteau} sample.

Whilst this process has an obvious statistical objectivity, it is not clear that
it has any basis in physics. However, using $R_{min}$ to denote the radial 
position of the innermost remaining point of any given ORC after the application
of this process, it turns out that $R_{min}$ is a very strong, but noisy, tracer of
the optical radius, $R_{opt}$, as defined, for example, by
Persic \& Salucci \cite{Persic1995}. 
The implication of these correlations, given their strength, is that the
computed $R_{min}$ has some physical significance - possibly as a transition
boundary between bulge-dominated and disc-dominated dynamics.
The details of these correlations, for all
the samples analysed, are given in Appendix \ref{RminRopt}.
\subsection{The representation of $\ln A$}
All the $\ln A$ frequency diagrams shown in this paper are obtained using the 
same bin-width ($\Delta \ln A \,=\,0.055$) and initial point ($\ln A\,=\,2.2$)
that were used in the original paper, Roscoe \cite{RoscoeA}, on this topic.
There are therefore no hidden degrees of freedom available to enhance the
signals being discussed.
\subsection{The absolute determination of TF zero points}
Tully-Fisher methods are central to our analysis and, as is well understood,
whilst TF calibrations provide an absolute determination of the gradient term, 
the value of the zero point is considered to depend on the value of $H$ assumed 
for the general calibration procedure.

In accordance with this, each sample is analysed by constraining our TF
gradients to lie within the quoted error bars of the absolute TF gradient
determinations made by the astronomers who compiled each of the various samples.
We then consider if, for each sample, a zero point can be determined which gives
rise to the significant peak structure of the hypothesis.
If the answer to this question is positive then, in practice, a method for the 
absolute determination of the zero points will have been demonstrated.
\subsection{A simple diagnostic guide for TF calibrations}
For ideal data, for
which no systematic bias of any kind exists, we should find a statistical equality
between Hubble magnitudes and Tully-Fisher magnitudes; that is, we should find
$M_{TF} \approx M_{Hubble}$ over the magnitude range of the sample whenever a sample is
quiet in the Hubble sense.
On the basis of this latter assumption, we apply the following simple
diagnostic guide as a means
of making quick - and usually reliable - judgements in our various analyses:
\begin{enumerate}
\item
Compute the regression model $M_{TF} = A M_{Hubble} + B$ on the sample for some assumed
value of $H$;
\item Adopt some standard reference range on Hubble magnitudes.
For example, we use $(-23.3,-18.2)$ which contains about 95\% of the Mathewson 
et al \cite{Mathewson1992} Hubble magnitudes with $H=85\,$km/sec/Mpc.
\item Use the regression model to compute the magnitude mapping
\begin{displaymath}
(-23.3,-18.2)_{Hubble}~ \rightarrow ~ (M_{min}, M_{max})_{TF},
\end{displaymath}
after rejecting outliers,
and use this mapping to make qualified judgements about the TF calibration relative to
the assumed value of $H$.
\end{enumerate}
\subsection{Other Essential Routine Issues}
There are two other essential, but routine, issues that are dealt with in 
the appendices. These are:
\begin{itemize}
\item Tully-Fisher methods are fundamental to this analysis and so it is
necessary to certain that Tully-Fisher scatter cannot wash out the signals being
claimed. This is dealt with in appendix \S\ref{Scatter.sec}.
\item There is always the possibility that the hypothesised effect is an
artifact. The various possibilities include the original process of measuring
ORCs, the methods of linewidth estimation and methods of folding.
These possibilities are discussed in \S\ref{Artifact.sec}.
\end{itemize}
\section{The Analysis of the Mathewson et al \cite{Mathewson1992} Sample}
\label{MFB.sec}
The Mathewson et al \cite{Mathewson1992} sample, like the 
Mathewson \& Ford \cite{Mathewson1996} sample, is drawn from an area of the
sky which Lynden-Bell et al \cite{Lynden-Bell} believe contain the Great 
Attractor (GA) and approximately one half of the 
Mathewson et al \cite{Mathewson1992} and Mathewson \& Ford \cite{Mathewson1996} samples lie 
within (Mathewson et al \cite{Mathewson1992} definition of) the GA region, 
$260^o < l < 360^o, -40^o < b < 45^o$.
In their figure 12, Mathewson et al \cite{Mathewson1992} use their 
Tully-Fisher calibration to show that
inside the GA region, the data exhibits a clear bias consistent with
some form of large-scale flow whilst, outside of the GA region, there is 
no such bias.
For this reason, we restrict applications of the magnitude mapping diagnostic on this
sample to the non-GA regions.
\subsection{The Mathewson et al calibration for MFB data}
Mathewson et al \cite{Mathewson1992} calibrated their Tully-Fisher relation 
against the Fornax
cluster (for which there is a very narrow redshift dispersion) on the basis of 
the assumption that Fornax is at $1340$km/sec (using $H=85$km/sec/Mpc).
Uniquely amongst the samples analysed here, their linewidth determinations were
made using an intuitive case-by-case {\lq eye-ball'} method (private
communication).
They obtained
\begin{equation}
M = - 8.18\,(\pm 0.12)\, \log V_{rot} - 2.86,
\label{eqn3}
\end{equation}
as their Malmquist bias corrected TF form.
In fact, Mathewson et al do not report error bars on their TF gradient, but do report
error bars of $\pm 0.26\,$mags for magnitude determinations in Fornax.
Our estimate of $\pm 0.12$ as the error bar for their gradient determinations is derived
from this.
Using their mean value gradient of $-8.18$ and the same zero point, with Mathewson's
assumed $H=85\,$km/sec/Mpc, we find that,
for the subsample {\it exterior} to the GA region, the diagnostic magnitude mapping
gives 
\begin{equation}
(-23.3,-18.2)_{H85} ~ \rightarrow ~ (-23.0, -18.1)_{TF}
\label{eqn3a}
\end{equation}
which indicates a possible discrepancy at the bright end between Hubble and Tully-Fisher
magnitudes in the non-GA region.

\begin{figure}
\noindent
\begin{minipage}[b]{\linewidth}
\resizebox{\hsize}{!}{\includegraphics{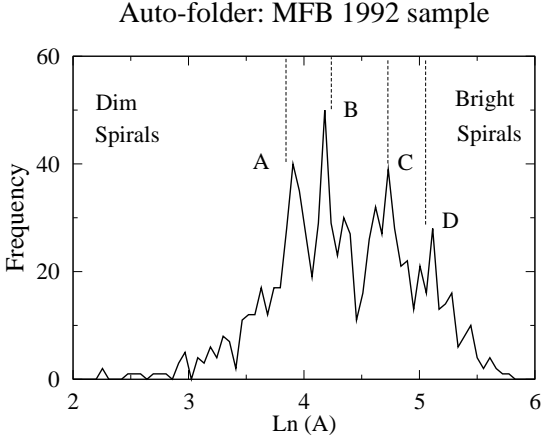}}
\end{minipage} 
\caption{$\ln A$ distribution for the Mathewson et al \cite{Mathewson1992}
sample with auto-folding and original Mathewson et al TF scaling; 
Vertical dotted lines indicate peak centres of Persic \& Salucci solution.
Bin width = 0.055}
\label{fig1}
\end{figure}

The $\ln A$ frequency distribution for the Mathewson et al sample with the 
calibration (\ref{eqn3}) (with Mathewson's quoted $-8.18$) generated by our own 
folding technique is shown in figure \ref{fig1}.
\begin{figure}
\noindent
\begin{minipage}[b]{\linewidth}
\resizebox{\hsize}{!}{\includegraphics{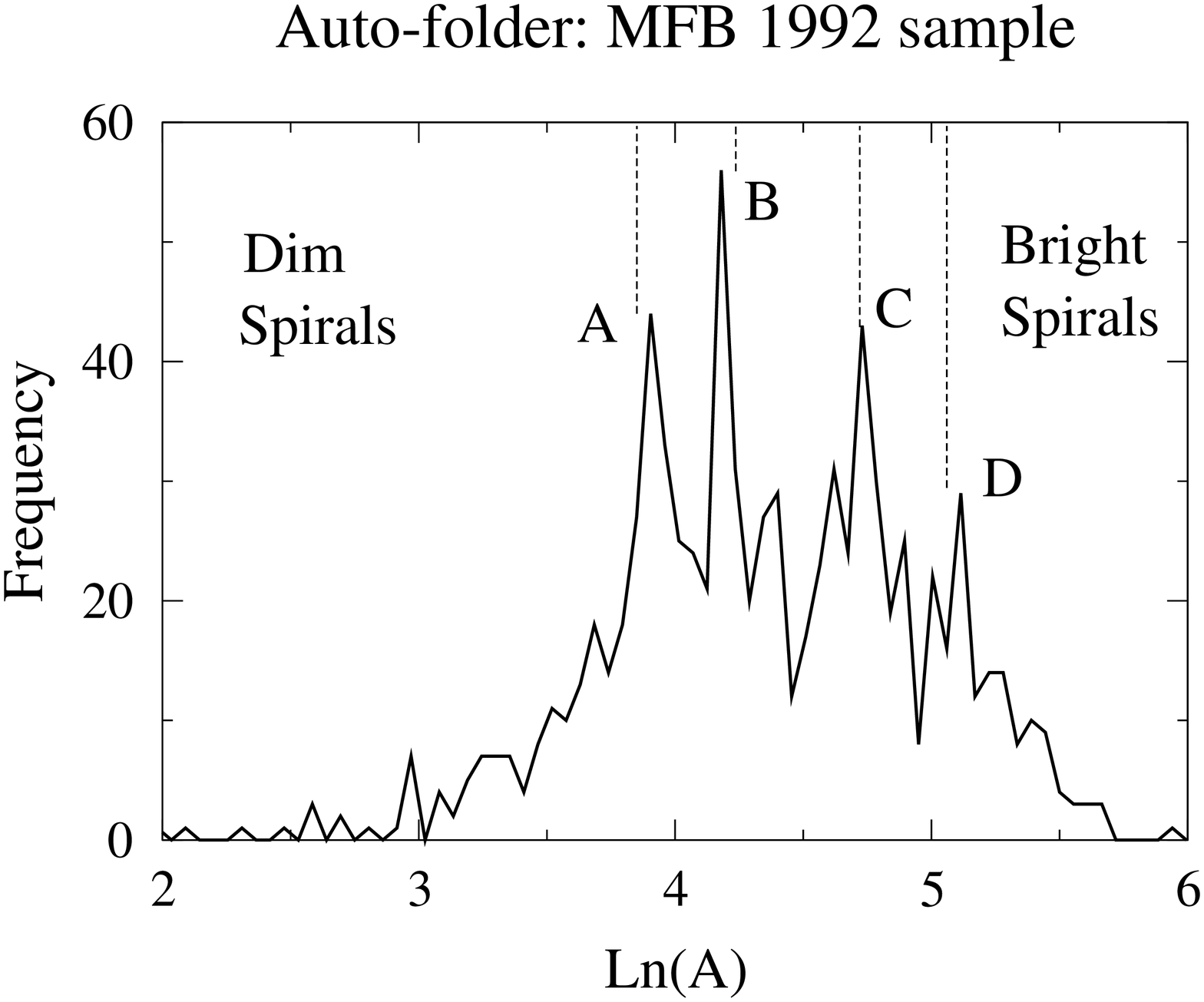}}
\end{minipage} 
\caption{$\ln A$ distribution for the Mathewson et al \cite{Mathewson1992}
sample with auto-folding and rescaled TF; 
Vertical dotted lines indicate peak centres of Persic \& Salucci solution.
Bin width = 0.055}
\label{fig1B}
\end{figure}
The vertical dotted lines in this latter figure mark the positions of the peaks 
$A,\,B,\,C,\,D$ of the Persic \& Salucci \cite{Persic1995} solution, in figure 
\ref{fig1A}. 
It can therefore 
be seen that the peak structure revealed by the Persic \& Salucci \cite{Persic1995} 
folding method is not an 
artifact of their method, but is an objective feature on the 
Mathewson et al \cite{Mathewson1992} sample.
\subsection{The effect of rescaling the TF relation within the error bars}
According to the diagnostic magnitude mapping, (\ref{eqn3a}), the reference range of
TF magnitudes generated by the scaling (\ref{eqn3}) is compressed and shifted to the dim end
relative to the Hubble magnitudes.
This suggests that a possible rescaling designed to expand the reference range of TF
magnitudes could be considered.
Such an expansion, consistent with Mathewson et al's original calibration, is obtained by 
{\it steepening} the TF gradient within the quoted error bars. 
We find that, if the peak structure is to be optimized, then a steepening of the gradient
(in the negative sense) must be accompanied by a reduction in the zero point.
The overall process is completed by an {\lq eye-ball'} iteration which can be described as 
follows:
\begin{enumerate}
\item Guess a new gradient value within the quoted error bar;
\item Adjust the zero point to maximise the peak structure for this gradient;
\item Perform the diagnostic magnitude mapping, adjusting $H$ as necessary to get the
best match at any given iteration;
\item Repeat until a satisfactory magnitude mapping is obtained.
\end{enumerate}
We find
\begin{equation}
M = - 8.30\, \log V_{rot} - 2.56,
\label{eqn3C}
\end{equation}
gives the magnitude mapping 
\begin{equation}
(-23.3,-18.2)_{H95} ~ \rightarrow ~ (-23.2, -18.2)_{TF}\,
\label{eqn3D}
\end{equation}
computed with $H=95\,$km/sec/Mpc.
There is now no significant magnitude bias or shifting.
The corresponding $\ln A$ frequency map is given in figure \ref{fig1B}, and we see
a small increase in signal strength on all four peaks.
In summary, the combined effects of the improved magnitude mapping, together 
with the increase in the 
strength of the $\ln A$ peak structure suggest that the recalibration 
(\ref{eqn3C}) is justified.
\subsection{The effect of excluding the edge-on galaxies}
In practice, we find that excluding the edge-on galaxies, which are
about 17\% of the sample, has the effect of producing only a pro-rata reduction
in the heights of the peaks in figure \ref{fig1}.
In other words, the {\it inclusion} of the edge-on galaxies in the present
analysis has a neutral effect upon it.
The reason for this result is almost certainly that the effects of
internal absorption are considerably greater on the interior parts of ORCs than
they are on the exterior parts, and
it is precisely these parts which are discarded in the data reduction
process described in \S\ref{Sec4} for the computation of $(\alpha,\,\ln A)$.
\section{The analysis of the Dale, Giovanelli, Haynes \& Uson sample}
\label{Sec.DGHU}
From our point of view, the value of the work of Dale et al \cite{Dale1997} et seq
lies in its provision
of a rotation curve sample which is completely independent in all of its
aspects of the Mathewson et al samples.

As a general comment, it is to be noted that part of the explicit
general programme of Giovanelli, Haynes et al has been to develop a highly 
accurate Tully-Fisher $I$-band {\lq template'} relation for the purpose of
establishing a cluster inertial frame out to $z \approx 0.06$.
Thus, Giovanelli et al \cite{Giovanelli} are able to quote the very tight 
error bars of $\pm 0.02$ for their TF gradient determinations.
To ensure such accuracy, 
much effort has been put into establishing a reliable linewidth
estimation technique.
This technique, which is described by 
Dale et al \cite{Dale1998}, is an algorithmic approach based 
on the estimate $V_{90\%}-V_{10\%}$ first introduced by 
Dressler \& Faber \cite{Dressler}.
However, it should be noted that this definition is most reliable when ORCs 
extend out to at least $R_{83}$, the optical radius as defined by Persic \&
Salucci \cite{Persic1995}.
The reason is that, as Dale et al \cite{Dale1998} point out, the
definition then recovers $V_{opt}$ in a reliable way and, as Persic \& Salucci
\cite{Persic1991,Persic1995} have shown, $V_{opt}$ provides a reasonable basis
for linewidth definitions.
To deal with those ORCs which were insufficiently extensive, Dale et al 
\cite{Dale1998}
introduced an elaborate procedure which involved the use of $N_{II}$ 
measurements to calculate a parameter, called a {\lq shape factor'}, for 
each ORC. This shape factor was subsequently used to extrapolate ORCs out to
$R_{83}$ and to correct raw $V_{90\%}-V_{10\%}$ linewidths as necessary.
\subsection{The Dale et al calibration of Tully-Fisher}
\begin{figure}
\noindent
\begin{minipage}[b]{\linewidth}
\resizebox{\hsize}{!}{\includegraphics{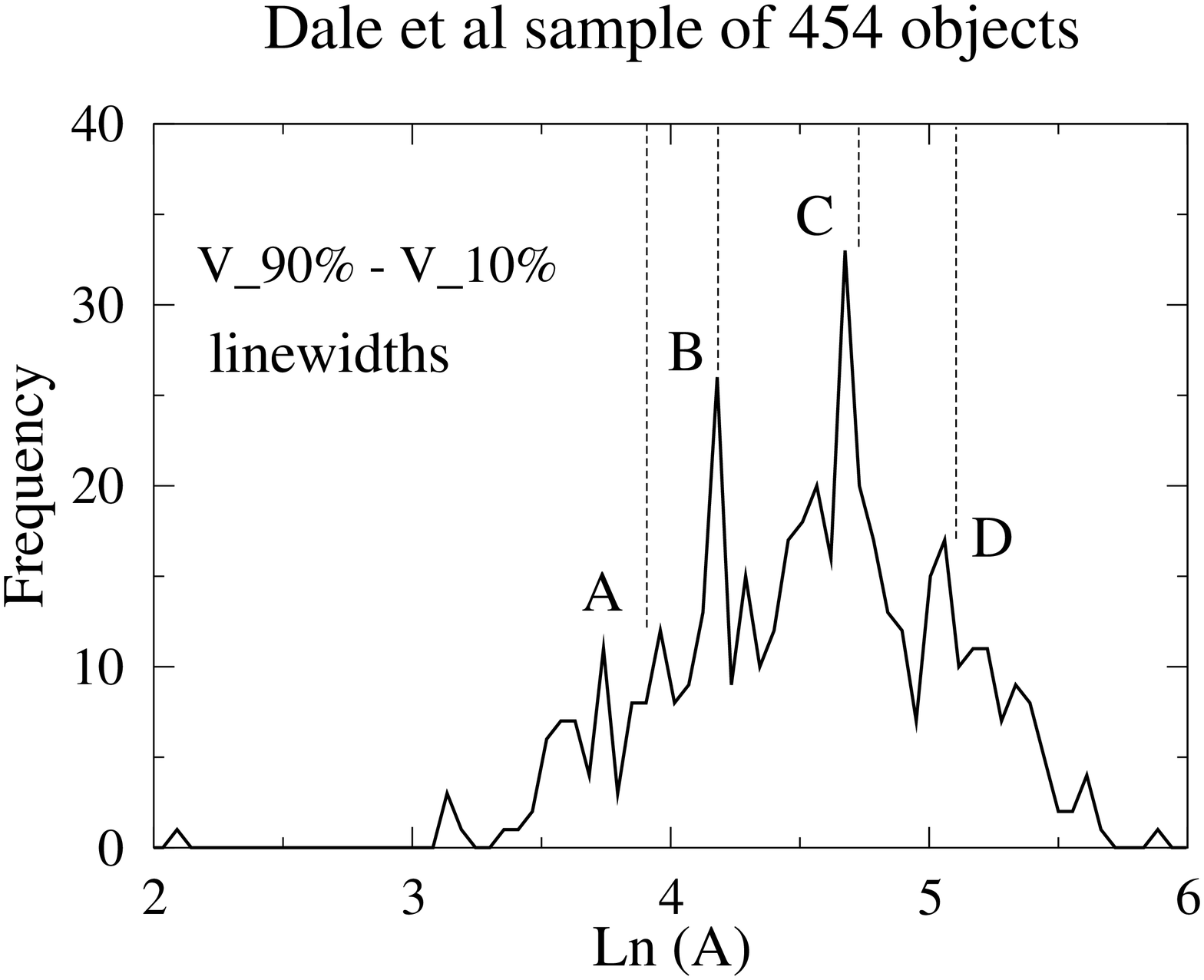}}
\end{minipage} 
\caption{The $\ln A$ distributions for the Dale et al
sample, with auto-folder and using the Dale et al TF calibration for
$H=95\,km/sec/Mpc$.
Vertical dotted lines indicate peak centres of
figure \ref{fig1}. Bin width = 0.055}
\label{fig2C}   
\end{figure}

The Dale et al \cite{Dale1998} calibration of the Tully-Fisher relation is given
by
\begin{equation}
M_{TF} = -7.68\,(\pm 0.02)\, \log V_{rot} - 4.12 + 5\,\log h
\label{eqnDale0}
\end{equation}
where $H = 100 h\,$km/sec/Mpc is undetermined.
The error bars on the gradient are so tight that, effectively, there is no
freedom to vary the estimated TF gradient.
However, the analysis of the Mathewson et al sample in \S\ref{MFB.sec} strongly 
suggest a value $H = 95\,$km/sec/Mpc so that $h = 0.95$ in (\ref{eqnDale0}).
Thus the suggested calibration is given by
\begin{equation}
M_{TF} = -7.68 \log V_{rot} - 4.23, \label{eqnDale1}
\end{equation}
and the corresponding $\ln A$ frequency diagram is given in figure
\ref{fig2C}. 
We see that the predicted peak structure is strongly reproduced.
Similarly, the diagnostic magnitude mapping is given by
\begin{displaymath}
(-23.3,-18.2)_{H95} \rightarrow (-23.2,-18.1)_{TF}
\end{displaymath}
which, apart from a very slight uniform shift to the dim, indicates the complete
absence of any systematic magnitude bias. 
\subsection{The effect of excluding the edge-on galaxies}
About 11\% of the Dale et al sample consists of edge-on
galaxies, and the effect of excluding these galaxies is the same as for the
previous two samples - that is, there is merely a pro-rata reduction in
the peak heights.
\section{The Analysis of the Mathewson \& Ford \cite{Mathewson1996} sample}
\label{MF.sec}
The Mathewson \& Ford \cite{Mathewson1996} sample, like the Mathewson et al 
\cite{Mathewson1992} sample, is also drawn from an area of the
sky which contains the, so-called, GA and approximately one half of the 
Mathewson \& Ford \cite{Mathewson1996} 
sample lies within the GA region, $260^o < l < 360^o, -40^o < b < 45^o$ but, as
reference to table \ref{Table3} shows, is an average of 70\% more distant than
the Mathewson et al \cite{Mathewson1992} sample and is therefore considerably 
less bright. 
\subsection{The Mathewson et al calibration for MF data}
\begin{figure}
\noindent
\begin{minipage}[b]{\linewidth}
\resizebox{\hsize}{!}{\includegraphics{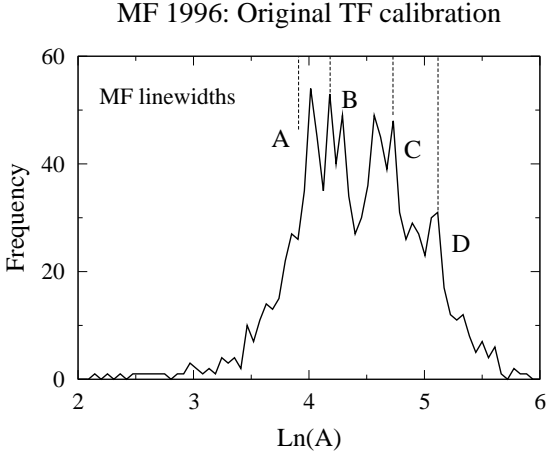}}
\end{minipage} 
\caption{The $\ln A$ distributions for the Mathewson \& Ford \cite{Mathewson1996} sample 
with auto-folding using  Mathewson linewidth estimates and original TF 
calibration.
Vertical dotted lines 
indicate peak centres of figure \ref{fig1}. Bin width = 0.055}
\label{fig2F}
\end{figure}

A direct application of the Mathewson et al \cite{Mathewson1992} TF calibration,
given at (\ref{eqn3}), to the Mathewson \& Ford \cite{Mathewson1996} sample
gives the $\ln A$ frequency diagram of figure \ref{fig2F}.
This figure shows that, whilst the $B$, $C$ and $D$ peaks are reproduced, 
the distribution is extremely noisy and must be considered poor.
Given the quality of the results obtained from the Mathewson et al
\cite{Mathewson1992} sample, and shown in figures \ref{fig1} and \ref{fig1B}, 
we can suppose that some kind of problem exists.

As a possible means of understanding what the problem might be, we consider the
subsample of Mathewson \& Ford \cite{Mathewson1996} data which is 
exterior to the GA region (discussed in \S\ref{MFB.sec}) and calculate how
the reference range of Hubble magnitudes maps into Tully-Fisher magnitudes
for this data.
Using the original Mathewson TF calibration (for which $H = 85\,$km/sec/Mpc), 
we find the diagnostic magnitude mapping for non-GA objects
\begin{equation}
(-23.3,-18.2)_{H85} ~ \rightarrow ~ (-23.0,-17.4)_{TF} 
\label{eqn5}
\end{equation}
which implies the existence
of a very strong systematic bias which shifts the whole  
Mathewson \& Ford \cite{Mathewson1996} sample to lower luminosities.
This is improved at the bright end when our modified Mathewson TF calibration
(\ref{eqn3C}) is used, but the dim-end bias remains essentially unchanged..
Given that the non-GA objects are not believed to be participating in any
large-scale flow,  there are two basic possibilities for explaining the 
mapping (\ref{eqn5}) which can be listed as
\begin{itemize}
\item The MF Hubble luminosities are very much overestimated at the dim end;
\item The MF Tully-Fisher luminosities are very much underestimated at the dim
end.
\end{itemize}
The first possibility seems unlikely since Mathewson \& Ford \cite{Mathewson1996} photometry is in the $I$ band for 
which the internal and external extinction mechanisms are well understood, and 
for which well-tested correction techniques exist and have been applied by 
Mathewson \& Ford \cite{Mathewson1996}.
The second possibility would necessarily have its source in the systematic
underestimation of dim-end optical linewidths for this relatively distant 
sample.
Since Mathewson \& Ford \cite{Mathewson1996} (and Mathewson et al \cite{Mathewson1992}) 
used an intuitive {\lq eyeball'} technique for linewidth
estimation, this latter possibility seems the most likely explanation.
On the basis of the working hypothesis that Mathewson \& Ford's linewdith
estimates are subject to a systematic bias relative to the Mathewson et al
\cite{Mathewson1992} linewidths, there are now two possible ways to proceed:
\begin{enumerate}
\item
We can either derive our own linewidth estimates directly from the
sample using some algorithmic technique;
\item Or we can take advantage of the idea that, where systematic linewidth 
bias exists, it can accounted for by a compensating recalibration of the 
Tully-Fisher relation.
\end{enumerate}
We consider both of these approaches in the following sections.
\subsection{TF based on $V_{opt}$ linewidths}
The first approach is based on the idea of generating our own
linewidth estimates.
The most simple algorithmic technique for linewidth estimation is that based on
$V_{90\%}-V_{10\%}$.
However, as discussed in \S\ref{Sec.DGHU}, linewidths based on this definition
can only reliably be used when ORCs extend out to at least to the optical 
radius, and Dale et al used an elaborate procedure to negotiate this problem for
those ORCs which did not.
This approach is not possible in the present case because
the required $N_{II}$ measurements are not available for Mathewson data.
But Dale et al \cite{Dale1998} also report that their definition of 
linewidth recovers Persic \& Salucci's $V_{opt}$ whenever ORCs are sufficiently
extensive.
This implies that using $V_{opt}$ as a tentative linewidth definition for 
Mathewson \& Ford data should allow a TF calibration very similar to that of
Dale et al's given at (\ref{eqnDale1}).

\begin{figure}
\noindent
\begin{minipage}[b]{\linewidth}
\resizebox{\hsize}{!}{\includegraphics{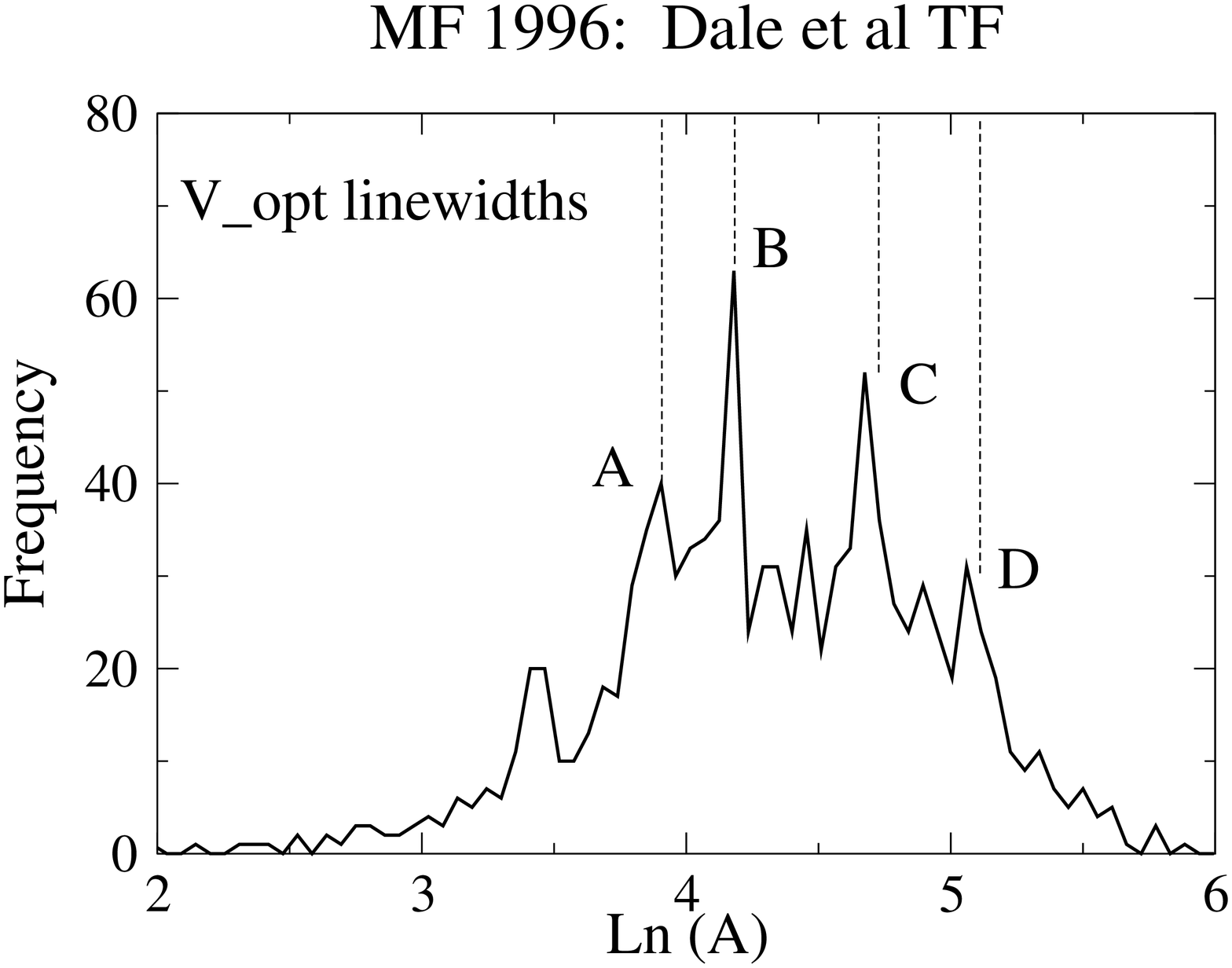}}
\end{minipage} 
\caption{The $\ln A$ distributions for the Mathewson \& Ford \cite{Mathewson1996} sample 
with auto-folding using $V_{opt}$ linewidths and Dale et al TF calibration with shifted zero point. 
Vertical dotted lines 
indicate peak centres of figure \ref{fig1}. Bin width = 0.055}
\label{fig2E}
\end{figure}

However, although $R_{83}$ (angular measure) estimates are provided with the 
Mathewson \& Ford 
\cite{Mathewson1996} sample, the corresponding $V_{opt}$ are not.
We circumvent this problem by using the power-law $V=A\,R^\alpha$ fitted to the
folded, but as yet unscaled, ORC to estimate $V_{opt}$ at $R_{83}$ 
(angular measure).
This process is possible prior to scaling simply because $\alpha$ values are 
independent of scaling.
With these linewidths, we find that the exact Dale et al calibration 
(\ref{eqnDale0}) with a zero point of $-4.08$,
\begin{displaymath}
M = -7.68 \log V_{rot} - 4.08,
\end{displaymath}
gives the $\ln A$ frequency diagram of figure \ref{fig2E}.
By contrast with figure \ref{fig2F}, we see that the whole $A,B,C,D$ peak
structure is now very well reproduced.
Interestingly, and as judged by the absence of any bias in magnitude mapping
\begin{displaymath}
(-23.3,-18.2)_{H85} ~ \rightarrow ~ (-23.2,-18.3)_{TF}, 
\end{displaymath}
this latter TF calibration corresponds more closely to $H = 85\,$km/sec/Mpc
for the non-GA Mathewson \& Ford \cite{Mathewson1996} subsample than the higher
value of $H = 95\,$km/sec/Mpc.
\subsection{Linear rescaling of TF for linewidth bias-correction}
\begin{figure}
\noindent
\begin{minipage}[b]{\linewidth}
\resizebox{\hsize}{!}{\includegraphics{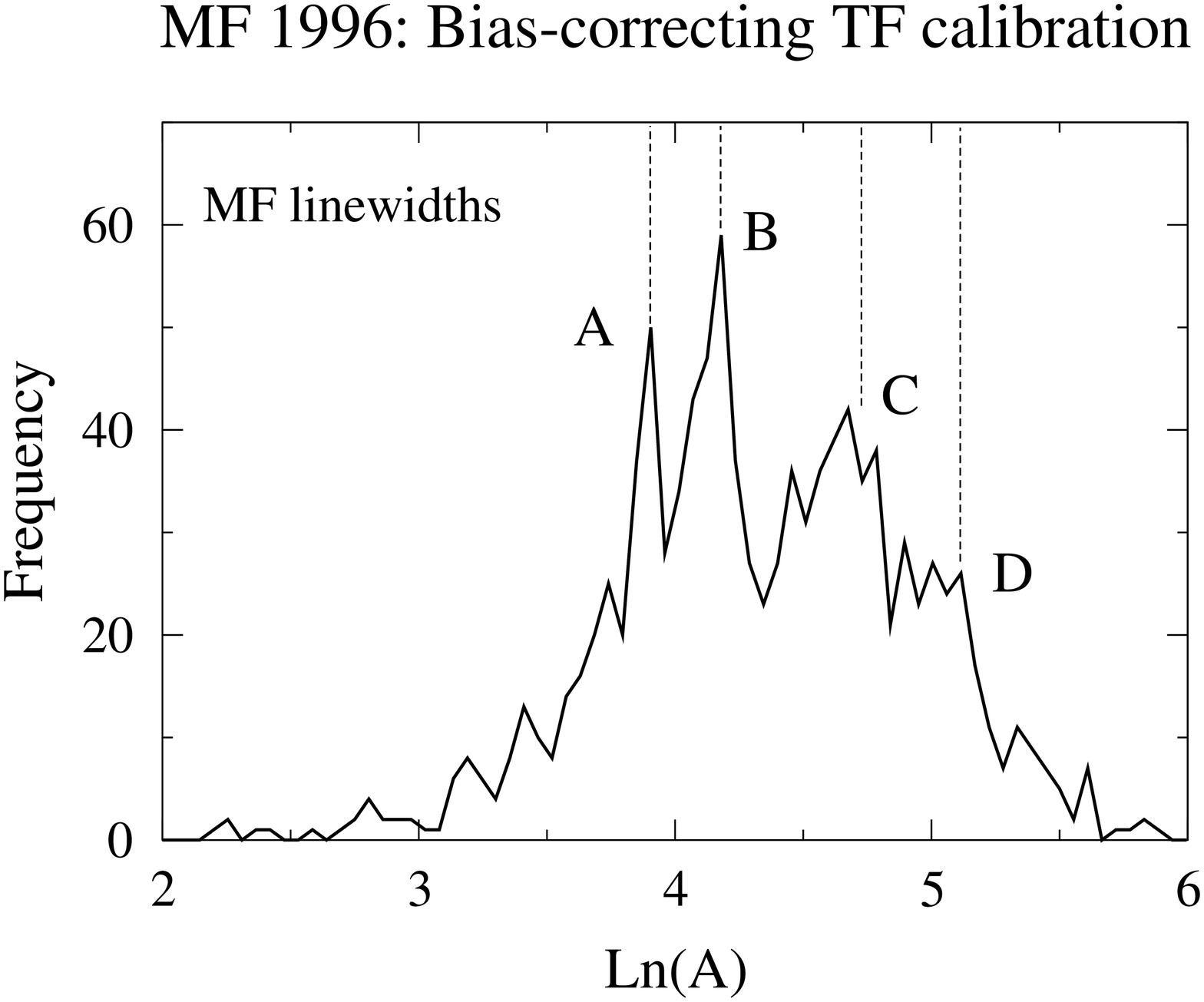}}
\end{minipage} 
\caption{The $\ln A$ distributions for the Mathewson \& Ford \cite{Mathewson1996} sample 
with auto-folding using the TF recalibration with linear rescaling. Vertical 
dotted lines 
indicate peak centres of figure \ref{fig1}. Bin width = 0.055}
\label{fig2B}
\end{figure}

The relative success of the $V_{opt}$ linewidth estimate provides further 
circumstantial evidence for the working hypothesis that the 
Mathewson \& Ford linewidths are, in fact, systematically biased.
We make the simplest possible assumption that such a bias can be accounted for
by a linear recalibration of the TF relation designed to ensure that the
diagnostic magnitude mapping is bias-free.
We find that the rescaled TF relation 
\begin{equation}
M_{TF} = -7.46 \log V_{rot} -  4.85,
\label{eqn4A} 
\end{equation}
gives the $H=88\,$km/sec/Mpc diagnostic magnitude mapping
\begin{equation}
(-23.3,-18.2)_{H88} ~\rightarrow ~(-23.3,-18.2)_{TF}, 
\label{eqn3b}
\end{equation}
for non-GA ORCs, which is perfect.
Corresponding to this improvement, we find that the recalibration (\ref{eqn4A})
gives the $\ln A$ frequency diagram of figure \ref{fig2B} which is likewise
a considerable qualitative improvement over figure \ref{fig2F}.
In particular,
the $A$ and $B$ peaks are now perfectly and strongly reproduced, whilst
the $C$ peak is clearly present, but noisy. The $D$ peak is virtually
non-existent.
\subsection{The effect of excluding the edge-on galaxies}
Only about 6\% of the Mathewson \& Ford (1995) sample consists of edge-on
galaxies, and the effect of excluding these galaxies is the same as for the
Mathewson et al \cite{Mathewson1992} sample - that is, there is merely a pro-rata reduction in
the peak heights.
\section{The analysis of the Courteau \cite{Courteau} sample}
\label{Courteau.sec}
As with the Dale et al sample, the value of the Courteau sample lies in 
its complete independence in all of its aspects of the Mathewson et al samples.

The Courteau \cite{Courteau} analysis was primarily designed to address 
the problem of linewidth definitions, with a view to obtaining a standardized 
objectively defined algorithmic definition.
Courteau considered several possibilities for linewidth definitions, and we
present results using his $V_{max}$ and $V_{2.2}$ definitions
(his estimated {\lq worst'} and {\lq best'} respectively).
Because his explicit concern was to make a comparative investigation of several
linewidth estimation techniques, less effort was expended in establishing
particularly accurate TF calibrations for each of the linewidth estimates
considered. 
For example, unlike Mathewson et at \cite{Mathewson1992} or Dale et al 
\cite{Dale1998}, Courteau did not calibrate
his Tully-Fisher relations on specially chosen low-redshift dispersion clusters,
but on his whole sample, which has a redshift dispersion in excess of
$10000\,$km/sec.
It follows that his calibrations are not likely to be as accurate as those of 
Mathewson et al \cite{Mathewson1992} and Giovanelli et al \cite{Giovanelli}.
Thus, Courteau's quoted error bars on gradient determinations for each of the
linewidth estimators considered are, typically, $\pm 0.2$,
which is to be compared with Giovanelli et al's \cite{Giovanelli} quoted error 
bar of $\pm 0.02$.
\subsection{The Courteau calibrations of Tully-Fisher for two linewidths}
\subsubsection{The $V_{max}$ linewidth TF calibration}
The Courteau \cite{Courteau} calibration of the Tully-Fisher relation for
his $V_{max}$ linewidths, using his $H=70\,$km/sec/Mpc, is given by
\begin{displaymath}
M_{TF} = - 6.09 \,(\pm 0.26)\,\log V_{max} - 7.22 \nonumber
\end{displaymath}
where the gradient error bar is as given by Courteau.
We subsequently find that the recalibration
\begin{displaymath}
M_{TF} = - 6.19 \log V_{max} -7.50
\end{displaymath}
gives the $\ln A$ frequency diagram of figure \ref{fig3}.
Except for the $A$ peak which is non-existent because of lack of data at the dim
end of the Courteau sample, we see that the peak structure of the hypothesis
is very well reproduced - although it is much noisier than the peak structures of
Mathewson et al and Dale et al shown in figures \ref{fig1B} and 
\ref{fig2C} respectively.
Note that the recalibrated gradient is comfortably within Courteau's quoted 
error bar.

It is also of interest to note that the net effect of the recalibration is to
increase the luminosity of the average galaxy by about $0.48$ magnitudes.
This is well within Courteau's quoted error bar for $V_{max}$ calibrations of
$0.55$ magnitudes and so there is no evidence to suggest that the recalibration
is significantly different from Courteau's original $V_{max}$ calibration
which assumed $H=70\,$km/sec/Mpc.
\subsubsection{The $V_{2.2}$ linewidth TF calibration}
The Courteau \cite{Courteau} calibration of the Tully-Fisher relation for
his $V_{2.2}$ linewidths, using his $H=70\,$km/sec/Mpc, is given by
\begin{displaymath}
M_{TF} = - 6.36\,(\pm 0.22) \,\log V_{2.2} -6.78 \,.
\end{displaymath}
We find that the recalibration
\begin{displaymath}
M_{TF} = - 6.55 \log V_{2.2} -7.15
\end{displaymath}
gives the $\ln A$ frequency diagram of figure \ref{fig7}.
Except for the $A$ peak which is non-existent because of lack of data at the dim
end of the Coruteau sample, we see that the peak structure of the hypothesis
is, again, reasonably well reproduced.
Note that the recalibrated gradient is within Courteau's quoted error bar.

This time, the net effect of the recalibration is to
increase the luminosity of the average galaxy by about $0.77$ magnitudes.
This is well outside Courteau's quoted error bar for $V_{2.2}$ calibrations of
$0.46$ magnitudes. Since the recalibrated gradient has not changed
significantly, this result is consistent with the idea that the recalibrated
zero point is significantly different from Courteau's original value for the
$V_{2.2}$ calibration.
\subsection{An oddness in the Courteau sample}
As a consistency check on the $V_{max}$ calibration, we note that the 
application of this 
calibration to the Courteau sample with $H = 70\,$km/sec/Mpc gives the 
magnitude mapping
\begin{displaymath}
(-23.3,-18.2)_{H70} \rightarrow (-23.0,-19.7)_{TF}.
\end{displaymath}
Thus, we see that Courteau's $V_{max}$ calibration compresses the magnitude range by about
35\%, with the dim end being too luminous by $1.5$ whole magnitudes.
It is easy to see that varying the Tully-Fisher zero point will simply have the
effect of shifting the given TF range $(-23.0,-19.7)_{TF}$ either up or down -
but it cannot expand the range to match the Hubble range.

Similarly, a consistency check on the $V_{2.2}$ calibration,  
with $H = 70\,$km/sec/Mpc gives the magnitude mapping
\begin{displaymath}
(-23.3,-18.2)_{Hubble} \rightarrow (-23.6,-19.6)_{TF}.
\end{displaymath}
The only way to get a match in either case is to vary both the gradient and 
the zero point in the Tully-Fisher relation.
We find that a good mapping can only be had with TF calibrations similar
to
\begin{displaymath}
M = - 9 \log_{10} V_{rot} -1.
\end{displaymath}
But the gradient here, $-9$, is well outside of the envelope of typical 
$R$-band TF gradients and can only be considered as extreme.
Given that Courteau's \cite{Courteau} study was primarily designed to
investigate algorithmic definitions of linewidths, and that he judged his
$V_{2.2}$ linewidths to be on a par with $H_I$ linewidths, it would seem that the
only likely explanation for the problem lies in the possibility that the sample
of chosen galaxies is not quiet in the Hubble sense.
\subsection{The effect of excluding the edge-on galaxies}
Less that 1\% of the Courteau sample consists of edge-on
galaxies, and so there is no discernible effect arising from their exclusion.
\begin{figure}[h]
\noindent
\begin{minipage}[b]{\linewidth}
%\resizebox{\hsize}{!}{\includegraphics{FIG5.eps}}
\resizebox{\hsize}{!}{\includegraphics{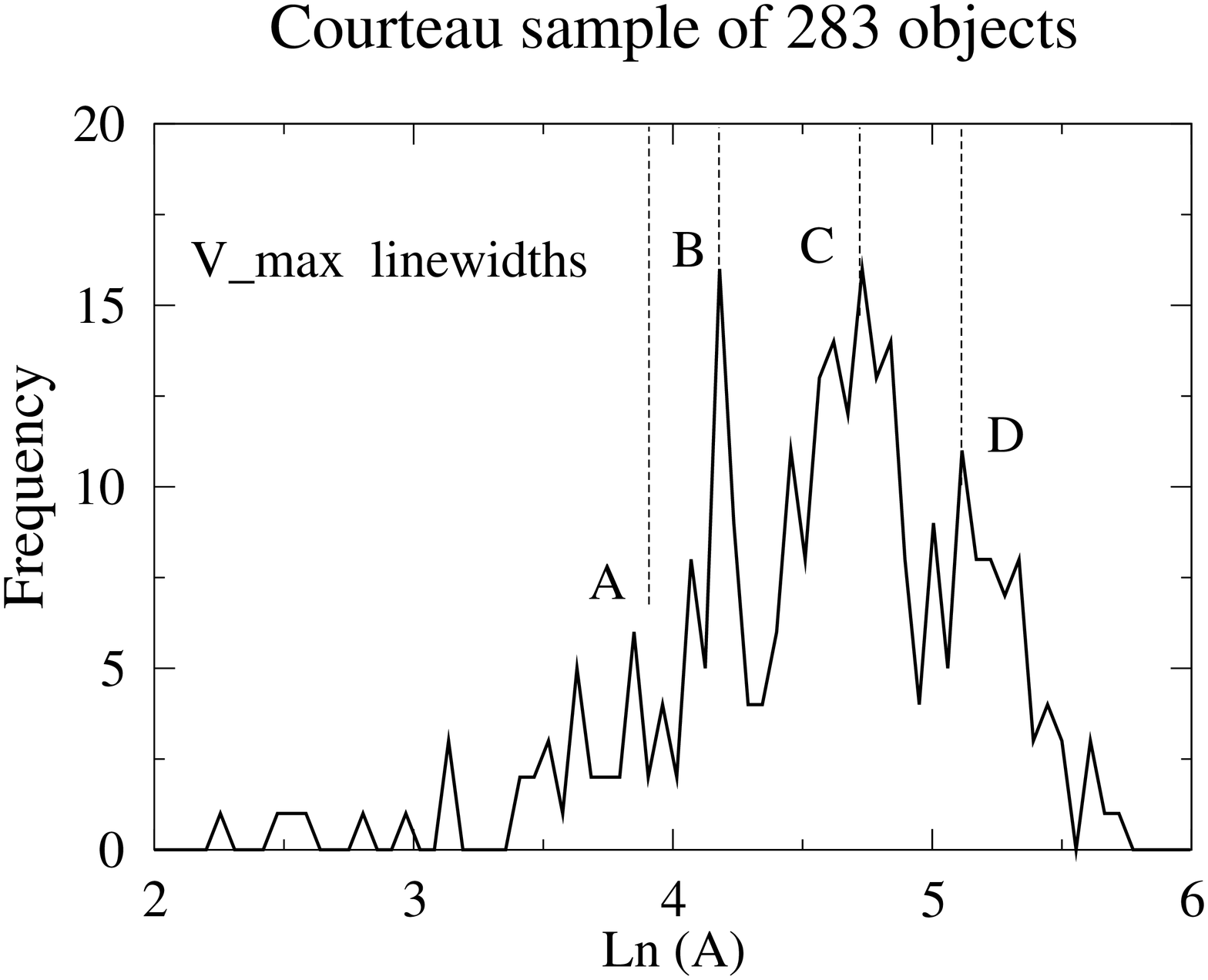}}
\end{minipage} 
\caption{The $\ln A$ distributions for the Courteau \cite{Courteau} sample 
using his $V_{max}$ linewidth definition and $H = 70\,$km/sec/Mpc. 
Vertical dotted lines indicate peak centres of
figure \ref{fig1}. Bin width = 0.055}
\label{fig3}
\end{figure}
\begin{figure}[h]
\noindent
\begin{minipage}[b]{\linewidth}
%\resizebox{\hsize}{!}{\includegraphics{FIG6.eps}}
\resizebox{\hsize}{!}{\includegraphics{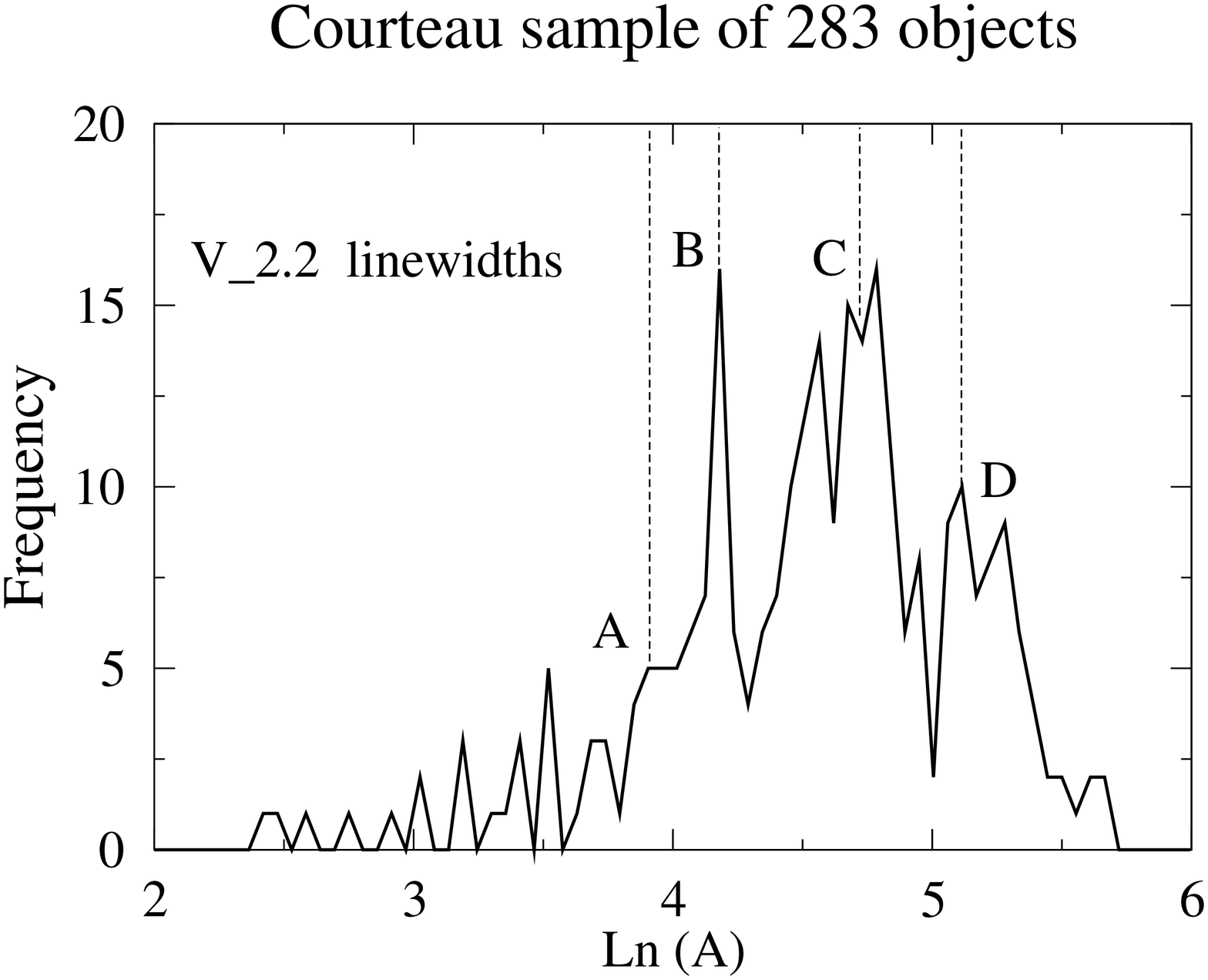}}
\end{minipage} 
\caption{The $\ln A$ distributions for the Courteau \cite{Courteau} sample using his
$V_{2.2}$ linewidth definition and $H = 70\,$km/sec/Mpc.
Vertical dotted lines indicate peak centres of
figure \ref{fig1}. Bin width = 0.055}
\label{fig7}
\end{figure}
\section{The statistical significance of the various analyses}
\label{Odds.sec}
We begin with a broad overview of the various analyses:
the results of the foregoing analyses, already encapsulated in figures
\ref{fig1}, \ref{fig1B}, \ref{fig2B}, \ref{fig2C}, \ref{fig7} and \ref{fig3}, are collected together for
convenient comparison in table \ref{Table6}.
The refined specific hypothesis to be defined and tested states, briefly, that 
strong peaks
in $\ln A$ frequency diagrams of rotation curve data processed in the way
described, should occur coincidently with the peaks $A,\,B,\,C$ and $D$ in figure
\ref{fig1}, which has been derived from our autofolder analysis of Mathewson et 
al (1992) data.
The exact bin-centre positions in which these latter peaks lie are given in the 
first row of table
\ref{Table6}, whilst remaining five rows give the peak centres for the remaining
figures.
The table makes it clear that the peak positions are essentially identical 
across the four samples.
In the following, we provide a standardized quantitative estimate of the 
statistical significance of these results.
\begin{table}
\begin{minipage}{2.0in}
\caption{Comparison of peak positions in $\ln A$ frequency diagrams for
MFB, MF, DGHU \& SC data}
\label{Table6}
\begin{tabular}{lllll}
\hline
Sample & A & B & C & D \\
\hline
MFB & 3.91 & 4.18 & 4.73 & 5.12 \\
${\rm MF}_{Vopt}$  & 3.91 & 4.18 & 4.68 & $5.06$ \\
${\rm MF}_{Vorig}$  & 3.91 & 4.18 & 4.68 & $-$ \\
DGHU &  $-$   & 4.18 & 4.68 & 5.06 \\
${\rm SC}_{Vmax}$ & $-$ & 4.18 & 4.73 & 5.12 \\
${\rm SC}_{V2.2}$ & $-$ & 4.18 & $4.73^*$ & $5.12$ \\
\hline
\multicolumn{5}{l}{$~^*$ indicates average over three points} \\
\multicolumn{5}{l}{$-$ indicates no significant peak} \\ 
\hline 
\end{tabular}
\end{minipage}
\end{table}
\subsection{A standardized methodology}
We noted, in \S\ref{Intro.sec}, that an extremely conservative upper-bound 
estimate of the probability of the peaks in figure \ref{fig1A}
occurring by chance alone, given the prior hypothesis raised on the
Rubin et al \cite{Rubin} data, was given in Roscoe \cite{RoscoeA} to be less than $10^{-7}$.
However, this estimate was derived using a crude ad-hoc methodology, and applied 
specifically to the Persic \& Salucci folding solution of the Mathewson et al 
(1992) data.
For the purpose of enabling statistical comparison across our whole analysis,  
we introduce a standardized methodology - which is already partly implemented -
and apply it where possible to each sample.
Specifically:
\begin{itemize}
\item
require that all samples to be tested are processed (folding etc), as far as is 
possible, in an identical fashion;
\item define the {\it null hypothesis} that all samples are drawn from the same
background distribution, and that this latter distribution is smooth - that is,
has no peak structure;
\item set up the specific hypothesis to be tested, and test it via a Monte-Carlo
simulation which randomly selects a very large number of samples from the 
hypothetical smooth distribution.
\end{itemize}
The first of these points is, of course, already implemented.
For the second point, experimentation shows that the final outcomes are 
insensitive to
any reasonable choice of {\lq smooth distribution'} for the null hypothesis;
we define it as the cubic spline envelope of the figure 
\ref{fig1} distribution shown in figure \ref{fig10}.
For the third point, the Mathewson et al \cite{Mathewson1992} sample must be tested against the
{\it original}
hypothesis raised on our analysis of the Rubin et al \cite{Rubin} sample, 
whilst our additional samples will
be tested against a refined hypothesis raised on the analysis of the 
Mathewson et al \cite{Mathewson1992} sample.
\subsection{Significance of the Mathewson et al \cite{Mathewson1992} results: 
Figure \ref{fig1}}
\label{Rubin_Hyp.sec}
In essence, we shall revise the original crude ad-hoc estimates of the 
significance of the
peaks arising from the Persic \& Salucci folding of Mathewson et al \cite{Mathewson1992} data,
shown in figure \ref{fig1A}, given the {\it original}
hypothesis raised on the Rubin et al \cite{Rubin} data described in \S\ref{Intro.sec}.
Briefly, this original hypothesis stated that, using Rubin scaling, then
$\ln A$ would lie within $\pm 0.15$ of integer or half-integer values -
specifically, the values 3.5, 4.0, 4.5 and 5.0.
The transformation of these Rubin-scaled intervals to the scaling used by 
Mathewson et al \cite{Mathewson1992} is  given in table \ref{Table8}.
\begin{figure}
\noindent
\begin{minipage}[b]{\linewidth}
\resizebox{\hsize}{!}{\includegraphics{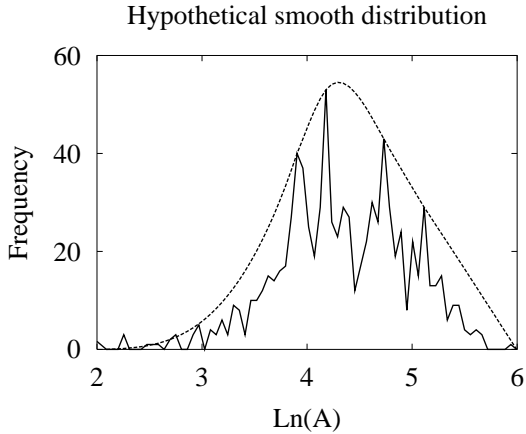}}
\end{minipage} 
\caption{Hypothetical smooth $\ln A$ distribution, defined as a cubic spline
envelope of actual MFB 1992 distribution.
Bin width = 0.055}
\label{fig10}
\end{figure}
\begin{table}
\caption{$\ln A$ interval transformation}
\label{Table8}
\begin{tabular}{lll}
\hline
Interval with & Interval with & Associated\\
RFT scale  & MFB scale  & peak\\
\hline 
$(\,3.35,\,3.65\,)$ & $(\,3.69,\,3.93\,)$ & $A$ \\
$(\,3.85,\,4.15\,)$ & $(\,4.10,\,4.34\,)$ & $B$ \\
$(\,4.35,\,4.65\,)$ & $(\,4.51,\,4.75\,)$ & $C$\\
$(\,4.85,\,5.15\,)$ & $(\,4.92,\,5.16\,)$ & $D$ \\
\hline
\end{tabular}
\end{table}
\begin{table}
\caption{Test of peak-structure in figure \ref{fig1}}
\label{Table9}
\begin{tabular}{lll}
\multicolumn{3}{c}{Peak frequencies in $10^6$ trials} \\
\hline
Interval with & Frequency at & Peak \\
MFB scale  & required & label \\
           & strength &   \\
\hline 
$(\,3.69,\,3.93\,)$ & $3513$ & $A$ \\
$(\,4.10,\,4.34\,)$ & $758$ & $B$ \\
$(\,4.51,\,4.75\,)$ & $18881$ & $C$ \\
$(\,4.92,\,5.16\,)$ & $66303$ & $D$ \\
\hline
\multicolumn{3}{c}{Probability of four sufficiently strong peaks} \\
\multicolumn{3}{c}{appearing in the same trial $\approx \,3 \times 10^{-9}$}
\\
\hline
\end{tabular}
\end{table}
The required probability estimate for the peaks of figure \ref{fig1} was 
obtained by
generating $10^6$ randomly selected samples of 866 measurements each (the number
of auto-foldable ORCs in the Mathewson et al \cite{Mathewson1992} sample) from the 
hypothetical smooth 
distribution of figure \ref{fig10}, and counting how many times peaks of the 
observed (or greater) sizes actually occur in the intervals specified in 
table \ref{Table8}.
The frequency at which peaks of the required size actually appeared is given in 
table \ref{Table9}.
On no occasion did more than two peaks at the required strengths appear in the
same trial; however, assuming independent probabilities, the observed 
frequencies allow us to estimate that the probability of four peaks of the
required strengths appearing simultaneously to be estimated at 
$3 \times 10^{-9}$.
\subsection{A refined hypothesis for the new samples}
\label{Sec.Refined_hyp}
The considerations of \S\ref{Rubin_Hyp.sec} allow us to refine the hypothesis
tested there (which arose from a consideration of just 12 Rubin et al \cite{Rubin}
objects) as follows: 
\hfill \break
\begin{it}
The distribution of $\ln A$, computed for folded ORCs according to the 
prescription of \S\ref{Sec4}, will show significant peak structure with
peaks centred on the $\ln A$ values $(3.91, 4.18, 4.73, 5.12)$, where 90\% 
confidence limits on the positions of these peaks, calculated in
detail in Appendix \ref{sec.ConfidenceLims}, are given {\rm approximately} by
$(3.87,3.98)$, $(4.15,4.21)$, $(4.70,4.76)$ and $(5.09,5.14)$ respectively.
\end{it} 
\begin{table*}
\caption{Test of peak-structure probabilities using the refined hypothesis
of \S\ref{Sec.Refined_hyp}}
\label{Table10A}
\begin{tabular}{lrrrrl}
\hline
 &\multicolumn{4}{c}{Peaks at observed strength in $10^6$ simulations}
 & Single trial \\
 & A & B & C & D & probability \\
 \hline
Fig\ref{fig2C} (Dale)& $-$    & ~~~~~11797 & ~~~~~3     & 23010 & $8\times10^{-10}$ \\
Fig\ref{fig2E} (MF $V_{opt}$)& 28106 &  ~~~~~131  & ~~~~~1204 & 70559 & $3\times10^{-10}$ \\
Fig\ref{fig2B} (MF $V_{orig}$)& 192 & ~~~~~1024 & ~~~~~93262 & $-$ & $2\times10^{-8}$ \\
Fig\ref{fig3} (SC $V_{max}$)& $-$ & ~~~~~44889 & ~~~~~8098 & 34653 & $1\times10^{-5}$ \\
Fig\ref{fig7} (SC $V_{2.2}$)& $-$ & ~~~~~44918 & ~~~~~17452 & 71252 & $6\times10^{-5}$ \\
\hline
\multicolumn{6}{c}{$-$ indicates non-existent peak.} \\
\hline
\end{tabular}
\end{table*}
\subsection{The significance of the Dale et al results: Figure \ref{fig2C}}
The error bars given by Giovanelli et al \cite{Giovanelli} for their TF gradient
estimate were so tight ($\pm 0.02$) that, effectively, there was no freedom
to vary this parameter.
Similarly, the $H=95\,$km/sec/Mpc value used to fix the TF zero point arose 
directly from optimizing the results of the Mathewson et al \cite{Mathewson1992} 
analysis given in \S\ref{MFB.sec}.
Effectively, therefore, the $\ln A$ frequency diagram of figure \ref{fig2C}
arose from a single trial and, accordingly, the associated probabilities can be
calculated directly from the figure on the basis of the modified hypothesis 
of \S\ref{Sec.Refined_hyp}.

The procedure is as before, except that each randomly selected sample now 
contains
454 measurements each - which is the number of auto-foldable ORCs in the 
Dale et al sample - and the results are shown in table
\ref{Table10A}.
Assuming independent probabilities, the observed 
frequencies allow us to estimate that the probability of the three peaks listed
appearing simultaneously at the observed magnitudes in figure \ref{fig2C}
to be estimated at $8 \times 10^{-10}$.
However, it is obvious from table \ref{Table10A} that most of the power in this latter
result arises from the extreme naure of peak $C$. But, even if the number of
peak $C$ occurrences equalled the number of peak $D$ occurrences, the 
single-trial probablility for this case would still be insignificant at
$6 \times 10^{-6}$.
\subsection{The significance of Mathewson \& Ford \cite{Mathewson1996} results: 
Figure \ref{fig2E}}
We now calculate the significance of the peak structure arising from our 
analysis of Mathewson \& Ford \cite{Mathewson1996} data - exhibited in figure 
\ref{fig2E} for the $V_{opt}$ linewidth results -
given the modified hypothesis of \S\ref{Sec.Refined_hyp}.
The procedure is as before, except that each randomly selected sample now 
contains
1085 measurements each - which is the number of auto-foldable ORCs in the 
Mathewson \& Ford \cite{Mathewson1996} sample - and the results are shown in table
\ref{Table10A}.

Assuming independent probabilities, the observed 
frequencies allow us to estimate that the single-trial probability of four 
peaks of the required strengths appearing simultaneously to be estimated at 
$3 \times 10^{-10}$.
However, in this particular case:
\begin{itemize}
\item We allowed ourselves the freedom to use an alternative
linewidth estimation method when the original linewidths of Mathewson \&
Ford \cite{Mathewson1996}, which gave rise to figure \ref{fig2F}, were 
inferred to be problematical;
\item Whilst, for the alternative method we used the Giovanelli et al gradient
value for the TF calibration, we allowed ourselves the freedom to search for a
new zero point. Discounting the fact that we were guided by the magnitude 
mapping
technique, it is a fair assessment to say that, typically, the search for a
suitable zero point required about five further independent trials.
\end{itemize}
To make a round figure, suppose then that the figure \ref{fig2E} results
required ten independent trials to obtain.
On this basis, we can use the single-trial probability and binomial statistics
to estimate the required probability at $3 \times 10^{-9}$. 
\subsection{The significance of the Courteau results: Figures \ref{fig7}}
The estimation of the probabilities attached to the Courteau results is a little
more involved than the other two case. Briefly, we argue as follows:
\begin{itemize}
\item We chose at the outset to consider only the $V_{max}$ and $V_{2.2}$
Courteau linewidth estimates. So there are two degrees of freedom arising here;
\item For each of the linewidth estimates, the error bars quoted by Courteau 
for the gradients are approximately $\pm 0.2$.
Given that we allow ourselves the freedom to vary the gradient within the quoted
error bars, this amounts to about three independent trials to optimize the
gradient value per linewidth estimate;
\item Finally, we allowed ourselves the freedom to determine the zero points for
each of the TF calibrations. Typically, each zero point determination 
required five trials for a given gradient determination.
\end{itemize}
So, for each linewidth estimate, we can reasonably say that fifteen independent
trials were required to determine the optimal TF calibration - making thirty
independent trials in all.
Given the single-trial probability of $1 \times 10^{-5}$ from table 
\ref{Table10A} for the $V_{max}$ linewidth analysis, we can use binomial
statistics to estimate the overall
probability of obtaining the figure \ref{fig3} results by chance alone at about
$3 \times 10^{-4}$.
\subsection{Summary of statistics}
A hypothesis was raised on the basis of a simple analysis of 12 Rubin et al
\cite{Rubin} galaxies; this hypothesis was tested against the results obtained from a 
sample of 866 Mathewson et al \cite{Mathewson1992} ORCs and it was confirmed 
with a
probability of $\approx 3 \times 10^{-9}$ against the obtained results arising
purely by chance.

The analysis of this larger sample of 866 ORCs allowed a refinement of the 
hypothesis, which
was subsequently tested against the results obtained from three new independent
samples of 
454 ORCs with $I$-band photometry from Dale et al \cite{Dale1997} et seq, 
1085 ORCs with $I$-band photometry from Mathewson \& Ford \cite{Mathewson1996} 
and 283 ORCs with $R$-band photometry from Courteau \cite{Courteau}.
The probabilities of the results obtained from each of these samples arising by
chance alone were estimated at $8 \times 10^{-10}$, $3 \times 10^{-9}$ and
$3 \times 10^{-4}$ respectively.
On the basis of these results, it is reasonable to say that the {\lq discrete
dynamical classes'} hypothesis for spiral discs has been verified at the
level of virtual certainty.
\section{A second generation of Tully-Fisher methods.}
\label{TF}
At a practical level, the foregoing analysis has  
implications concerning the general nature of Tully-Fisher methods which we
shall expand upon in this section.
These can be summarized as:
\begin{enumerate}
\item The discrete dynamical states phenomonology provides an absolute standard
whereby Tully-Fisher methods can be absolutely calibrated on a large enough
sample;
\item The Tully-Fisher method is shown to have two equivalent formulations, one
of which is the familar one; 
\item The Tully-Fisher relation is shown to be augmented by a corresponding
relation which defines {\it where} on an ORC a linewidth should be measured.
Thus, a means is provided whereby Tully-Fisher calibrations {\it and} linewidth
determinations can, in principle, be algorithmically, absolutely and 
simultaneously determined from any given large sample of ORCs;
\item The concept of {\lq linewidth'} is shown to be not absolute - that is,
linewidth can be defined with virtually unlimited freedom. 
But, once defined, the
form of the corresponding Tully-Fisher calibration becomes fixed.
\end{enumerate}
\subsection{The essential background}
\label{Luminosity.sec}
\begin{figure}
\noindent
\begin{minipage}[b]{\linewidth}
\resizebox{\hsize}{!}{\includegraphics{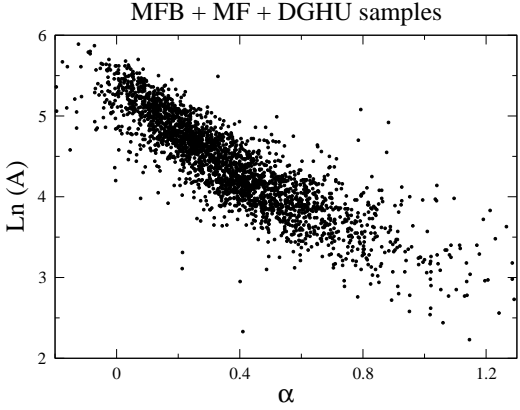}}
%\resizebox{\hsize}{!}{\includegraphics{XFIG15.eps}}
\end{minipage} 
\caption{Plot of $(\ln A,\,\alpha)$ for 2405 galaxies}
\label{FIG13A}
\end{figure}

The consequences listed above rest upon the assumption that the power-law 
analysis can be shown to provide a high-quality resolution of ORC data
in that part of the disc where the discrete dynamical classes phenomonology
manifests itself.
To demonstrate this, the first thing of significance to be considered is 
the $(\alpha,\,\ln A)$ plot,
given in figure \ref{FIG13A} for the combined Mathewson et al
\cite{Mathewson1992}, 
Mathewson \& Ford \cite{Mathewson1996}, Dale et al \cite{Dale1997} et seq 
samples of 2405 objects having $I$-band photometry and reliably
foldable ORCs.
\begin{table}
\caption{$\ln A$ model}
\label{Table7}
\begin{tabular}{lrr}
\hline
Variable & Coeff & t ratio \\
\hline 
${\rm const}$ & $-1.596$ & $-11$ \\
$M$ &$-.316$ &  $-46$ \\
$S$ & $-0.0002$ & $-2$ \\
$\alpha$ & 7.614 & 28 \\
$\alpha \,M$ & 0.474 & 34 \\
$\alpha \,S$ & 0.0050 & 18 \\
\hline
\multicolumn{3}{c}{ N=1951;  $R^2 = 92\%$}
\end{tabular}
\end{table}
We now make a detailed analysis of the 1951 $(\alpha,\,\ln A)$ points in 
figure \ref{FIG13A} which are associated with the specific subset of the
Mathewson et al 1992 + Mathewson \& Ford 1995 ORCs.
In the context of the model
\begin{equation}
{ V \over V_0 } = \left({ R \over R_0}\right)^\alpha~~\longrightarrow~~ 
A = {V_0 \over R_0^{\,\alpha}}\,,
\label{eqn6}
\end{equation}
we find
\begin{eqnarray}
\ln A &=& \ln\, V_0 - \alpha\,\ln\,R_0 \nonumber \\ 
      &=& -1.596 - 0.316 \,M   +7.614\,\alpha +0.474\,\alpha\,M \nonumber \\
    &~& + 0.0050\,\alpha\,S  \label{eqn7A} 
\end{eqnarray}
for the two $I$-band Mathewson samples.
Here $M$ is absolute magnitude, and
$S$ is surface brightness defined as average solar
luminosities per square parsec for the whole area inside the optical radius,
$R_{83}$, as defined by, for example, Persic \& Salucci \cite{Persic1995}.
This particular model was obtained by using the original Mathewson linewidths for the
Mathewson et al \cite{Mathewson1992} sample, and the $V_{opt}$ linewidths for
the Mathewson \& Ford \cite{Mathewson1996} sample, and by rejecting all 
$3 \,\sigma$ outliers (about 4\% of the total),
and it accounts for about 92\% of the variation in figure \ref{FIG13A}.
The detailed statistics are given in table \ref{Table7} and it is
clear that, except for surface brightness, $S$, all of the chosen predictors 
have an extremely powerful effect in the model. 
\subsection{Extraction of models for $\ln R_0$ and $\ln V_0$}
Introducing the {\lq discrete dynamical states'} phenomonology into the
discussion, we now note that the model can be decomposed 
into separate expressions for $(\ln R_0, \ln V_0)$ according to
\begin{eqnarray}
\ln A_i &=& \ln V_0 - \alpha \, \ln R_0,~~ i = 1,2,3,4 \nonumber \\
\ln  V_0 &=& -1.596 - 0.316 \,M  + \Gamma \label{7B} \\
\ln R_0 &=& -7.614\, -0.474\,M -0.0050\,S + {\Gamma \over \alpha}
\nonumber
\end{eqnarray}
where $\ln A_i \, = \,3.91,\,4.18,\,4.73,\,5.12$ and denotes the positions of 
the $A,B,C,D$ peaks, and $\Gamma$ is an arbitrary function (or parameter)
arising from the decomposition of $\ln A_i$.
It is easily verified that (\ref{eqn6}) is invariant with respect to an
arbitrary choice of this $\Gamma$.

\begin{center}
\begin{figure}[h]
\noindent
%\begin{minipage}[h]{\linewidth}
\resizebox{\hsize}{!}{\includegraphics{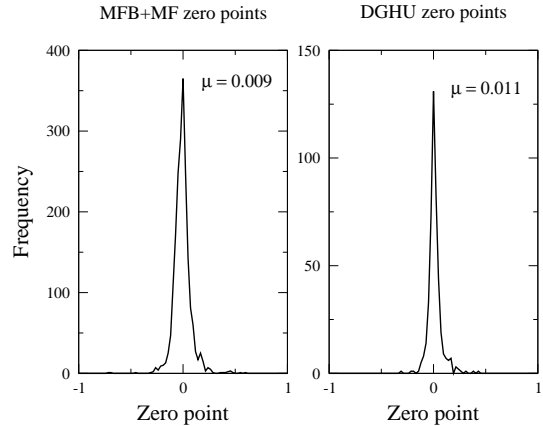}}
%\end{minipage} 
\caption{Plot of zero point for 1951 MFB+MF galaxies and 454 DGHU galaxies}
\label{FIG15}
\end{figure}
\end{center}

An effective visual verification of the fit of model (\ref{7B}) is obtained by
noting that if we use the definitions of (\ref{7B}) in the dimensionless form
of (\ref{eqn6}), and then regress
$\ln (V / V_0)$ on $\ln ( R / R_0)$, then we should find a null  zero point for 
each ORC - except for statistical scatter.
Because of the invariance of (\ref{eqn6}) with respect to $\Gamma$,
the results of this regression will be independent of any chosen $\Gamma$,
and so we can set it to zero for this particular exercise.
Figure \ref{FIG15} (left) gives the frequency diagram for the actual zero
points computed for the combined Mathewson et al (1992, 1995) samples from 
which the model (\ref{eqn7A}) was derived, whilst a wholly
independent test of the model is given by figure \ref{FIG15} (right) which gives the 
frequency diagram for the zero points derived from the Dale et al 
\cite{Dale1997} et seq 
sample using the model (\ref{eqn7A}).
It is clear that there is absolutely no evidence to support
the idea that these zero points are significantly different from the null
position - so that 
(\ref{eqn6}) with (\ref{7B}) can be considered to give a very effective
resolution of ORC data in the disc.
\subsection{Implications 1: The classical Tully-Fisher relation }
\subsubsection{Case $\Gamma = 0$: Retrieval of the standard formulation}
\label{Gamma=0}
We begin by noting that
a rearrangement of the second equation of (\ref{7B}), converting 
$\ln V_0$ to $\log\, V_0$, and setting $\Gamma = 0$ gives 
\begin{equation}
M\, = \, -7.29\,\log V_0 \,-\,5.05.\label{eqn2}
\end{equation}
Interpreting the characteristic scaling velocity $V_0$ as the disc 
{\lq rotation velocity'}, we can immediately recognize this relation as 
similar to a typical $I$-band Tully-Fisher calibration - but with a gradient 
on the low side.
However, this latter relation was derived by combining two
samples having different linewidth characteristics to make a specific point about the general analysis.
Consequently, we might
expect to find more typical gradients by analysing the four samples individually.
In fact, we find:
\begin{eqnarray}
M_I\, &=& \, -7.72\,\log V_0 \,-\,3.79 \nonumber
\\
M_I\, &=& \, -7.85\,\log V_0 \,-\,3.54 \nonumber
\\
M_I\, &=& \, -7.86\,\log V_0 \,-\,3.71 \nonumber
\\
M_R\, &=& \, -6.56\,\log V_0 \,-\,6.48 \nonumber
\end{eqnarray}
respectively for the Mathewson et al \cite{Mathewson1992}, Mathewson \& Ford
\cite{Mathewson1996}, Dale et al \cite{Dale1997} and Courteau \cite{Courteau}
samples - the latter using Courteau's $V_{max}$ linewidth definition as an
example.
We see that the three $I$-band gradients are now perfectly typical $I$-band TF gradients,
whilst the single $R$-band gradient is now a perfectly typical $R$-band TF gradient.
Similarly, for each case, the zero points are typical for 
$H \approx 85\,$km/sec/Mpc.
We can therefore reasonably conclude that the second equation of (\ref{7B})
with $\Gamma = 0$ recovers the standard formulation of the Tully-Fisher
relation.
  
It now follows directly that the third 
equation of (\ref{7B}), for the characteristic scaling radius $R_{0}$, 
effectively defines the position on an ORC at which the rotation velocity, $V_0$, is to 
be measured. That is, for the example given, the $\Gamma = 0$ case corresponds to 
measuring a characteristic rotation velocity $\ln  V_0 = -1.596 - 0.316 \,M$ at 
a characteristic radial position $\ln R_0 = -7.614\, -0.474\,M -0.0050\,S$.
\subsubsection{Case $\Gamma \neq 0$: Arbitrariness in calibration procedures 
for TF}
\label{GammaNEQ0}
Given the $\Gamma = 0$ case, it is now easy to see that the $\Gamma \neq 0$ case
corresponds 
to measuring a characteristic
rotation velocity, $V \equiv \exp(\Gamma)\,V_0$, at the radial position 
$R \equiv \exp(\Gamma /\alpha)\,R_0$.
Thus, according to the present considerations, there is an arbitrariness in 
calibrating Tully-Fisher relations which
consists in the freedom to choose which point on an ORC to use as the
characteristic radius for the measurement of the characteristic rotation 
velocity.
Also, since $\Gamma$ is an arbitrary {\it function} - and not just a constant -
this freedom of choice can extend to, for example, a procedure whereby the place
at which the rotation velocity is defined can be luminosity dependent.

This has significant consequences since, as mentioned in \S\ref{LWDs},
there are several non-equivalent heuristically defined methods which authors 
use to estimate optical 
linewidths and there is, as yet, no concensus about which method is {\lq best'}.
But, according to the above comments, it is clear that the {\lq best'} such 
method is the one which most consistently selects the radial positions 
$R \equiv \exp(\Gamma /\alpha)\,R_0$ across the sample under consideration, for
some $\Gamma$.
\subsection{Implications 2: Non-classical Tully-Fisher formulations}
\label{Impl2}
The foregoing considerations give rise to distinct Tully-Fisher formulations
neither of which is logically prior to the other.
They are equivalent in the sense of being equally applicable to the problem of
scaling.
\subsubsection{The quasi-classical formulation}
If we choose the $\Gamma = 0$ calibration, then (\ref{7B}) can be expressed
as:
\begin{eqnarray}
\ln V_0 &=& -1.596 - 0.316 \,M  \nonumber \\
\ln R_0 &=&- (1.596 + 0.316 \,M  + \ln A_i) / \alpha\, 
\label{eqnTF1} \\
i &=& 1,2,3,4.  \nonumber
\end{eqnarray}
As we saw in \S\ref{Gamma=0}, the first of these equations provides a typical 
$I$-band
Tully-Fisher calibration whilst the second effectively says where on an ORC
the linewidth, $V_0$, is to be measured. 
Note that, in this formulation, $\ln V_0$ varies only with $M$ as in any
conventional Tully-Fisher calibration, but $R_0$ varies with $M$ and the
dynamically measured quantities $(\alpha,\,\ln A)$.
\subsubsection{The non-classical formulation}
Equivalently, with the same $\Gamma = 0$ calibration, (\ref{7B}) can be 
expressed as
\begin{eqnarray}
\ln V_0 &=& - \alpha \,(7.338\, +0.461\,M +0.00449\,S) + \ln A_i\nonumber \\ 
i &=& 1,2,3,4 \nonumber \\
\ln R_0 &=& - ( 7.338\, +0.461\,M +0.00449\,S ).\nonumber
\end{eqnarray}
In this formulation, we see that the classical situation is reversed. 
Now $\ln V_0$ varies with $M$ and the dynamically measured quantities
$(\alpha,\,\ln A)$. Consequently, unlike in the classical formulation, both 
gradient and
zero point will vary across the luminosity/dynamical range of a sample.
By contrast, it is $R_0$ which varies with just luminosity 
properties, $M$ and $S$.
\subsection{Comments on general calibration procedures}
For the sake of explicitness, we shall use (\ref{eqnTF1}) to outline a suggested
algorithmic approach to the calibration procedure for a sufficiently large
sample.
The approach is non-trivial and has yet to be developed into a fully
working method:
\begin{enumerate}
\item Use the model (\ref{eqnTF1}), or something similar, as first guess at the
final calibration;
\item Fold an unscaled ORC and determine its $\alpha$ - which is a 
scale-independent quantity;
\item Estimate which $\ln A_i$ is appropriate for the ORC using the 
$(\alpha:\ln A)$ correlations implicit in figure \ref{FIG13A}; 
\item Iterate on the two equations of (\ref{eqnTF1}) and the power-law
$V = A\,R^\alpha$ fitted to the ORC, to find $(R_0,V_0,M)$ for the galaxy
concerned. The individual ORC can now be scaled;
\item Complete steps 2,3 and 4 for the whole sample and then:
\begin{enumerate}
\item calculate the new $\ln A$ frequency diagram and compare with some template 
distribution such as that of figure \ref{fig1};
\item
check if the calibration which comes {\it out} of the
model (in the manner of (\ref{eqn7A})) is the same as the
model which went in at step 1 - to within tolerance;
\end{enumerate}
If both these tests are satisfied, 
then accept the current calibration.
If not, adjust the calibration, and repeat steps 2, 3, 4 and 5 until converged.
\end{enumerate}
\section{Major Stability issues}
\label{Stability}
Apart from the importance of good Tully-Fisher calibrations and linewidth
determinations, the successful
extraction of the discrete dynamical states phenomonology from ORC data is also 
critically
dependent on the quality of the folding process and the algorithmic computation of
$\ln A$.
We consider these in turn.
\subsection{The quality of folding}
In the literature, the folding of ORCs is most commonly described in the context
of estimating linewidths for TF applications and, in such circumstances, the folding 
process generally amounts to determining an estimate of $V_{sys}$, the velocity 
equivalent of the systematic redshift. 
So far as we are aware, Persic \& Salucci \cite{Persic1995} were the first to address
the idea 
that ORCs folded for TF purposes were not necessarily folded with sufficient accuracy 
for the purpose of studying the interior dynamics of galaxy discs.
It was for this reason that they provided their own folding solutions for the
Mathewson et al \cite{Mathewson1992} sample.
Possibly their main innovations to the folding process were, firstly relaxing the 
assumption that the dynamic centre of a spiral coincided with its optical centre. 
Hence, their folding
process became a two-parameter problem - to determine $V_{sys}$ and the angular offset
between the dynamical and optical centres, $\Delta \Phi$ say.
Secondly, they introduced the pre-folding data filtering technique described in
\S\ref{PFDF} which has the effect of discarding, typically, the least accurate 40\% of 
velocity determinations on an given ORC.
They then applied a labour-intensive {\lq by-eye'} approach to the folding problem.
The folding algorithm developed for the present study (Roscoe \cite{RoscoeC})
followed Persic \& Salucci in utilizing both of these refinements.

\begin{center}
\begin{figure}
\noindent
%\begin{minipage}[h]{\linewidth}
\resizebox{\hsize}{!}{\includegraphics{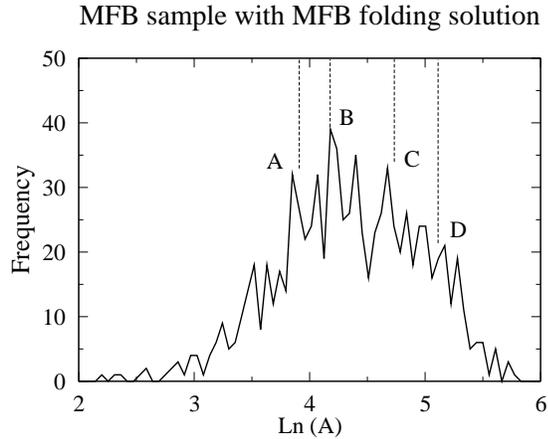}}
%\end{minipage} 
\caption{$\ln A$ distribution for the Mathewson et al \cite{Mathewson1992}
sample with MFB folding and original Mathewson et al TF scaling; 
Vertical dotted lines indicate peak centres of Persic \& Salucci solution.
Bin width = 0.055}
\label{FIG16}
\end{figure}
\end{center}

To illustrate the effect of folding without these refinements, figure \ref{FIG16} 
shows the $\ln A$ frequency diagram
computed for the Mathewson et al sample \cite{Mathewson1992}, but using Mathewson et
al's own folding solution.
This is to be compared directly with figures \ref{fig1A} and \ref{fig1}.
It is quite clear that the discrete states phenomonology is heavily obscured in 
figure \ref{FIG16}.
\subsection{The quality of $\ln A$ determinations}
The quality of $\ln A$ determinations is equally critical to the extraction of the
discrete states phenomonology.
Specifically, $\ln A$ is computed according to the {\lq black-box'} algorithm of 
\S\ref{TheAlgorithm}
which has the effect of, on average, discarding approximately the {\it inner} 10\% of 
any given ORC.
The value of $\ln A$ is then computed directly on the remaining outer segment of the
ORC.

\begin{center}
\begin{figure}
\noindent
%\begin{minipage}[h]{\linewidth}
\resizebox{\hsize}{!}{\includegraphics{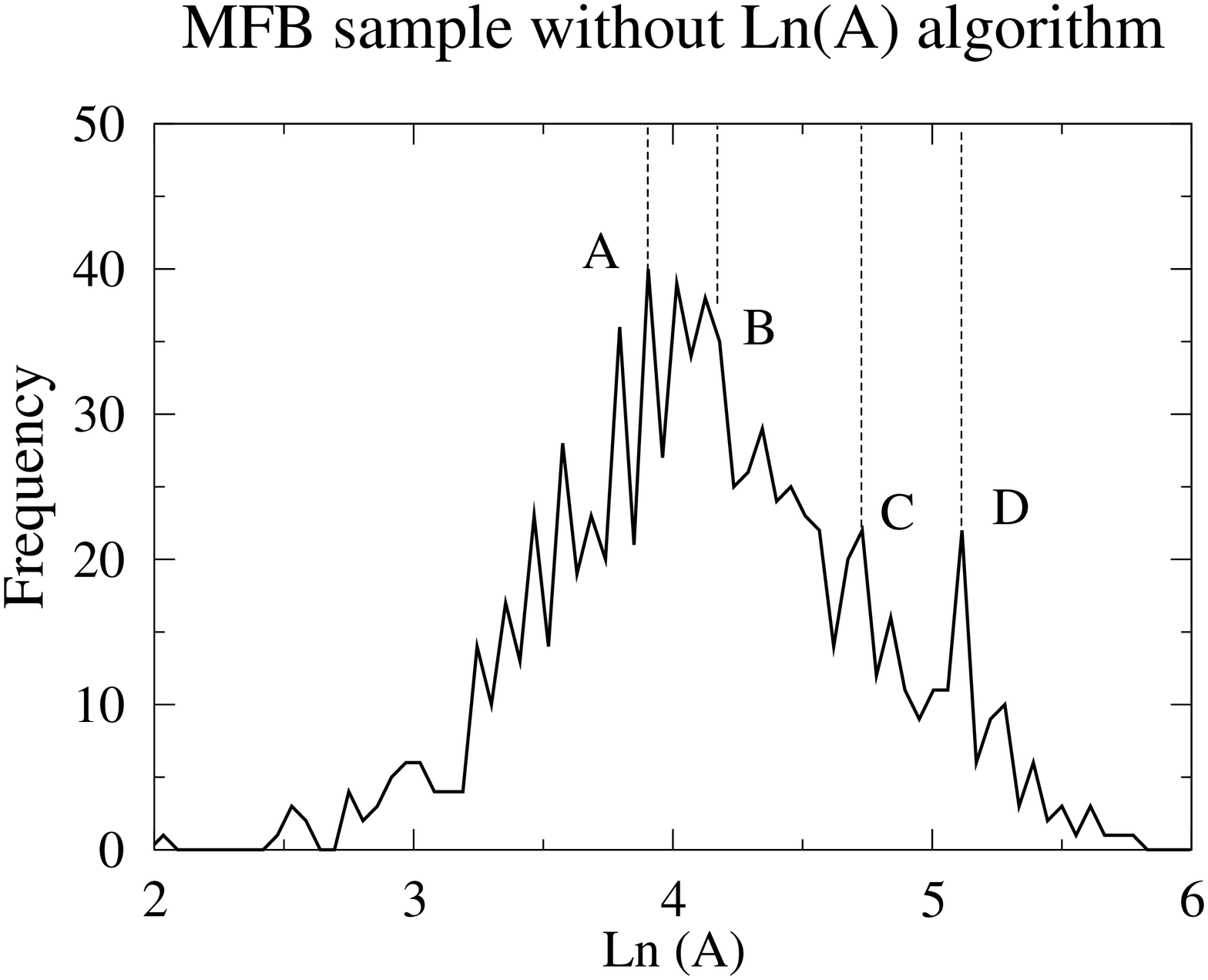}}
%\end{minipage} 
\caption{$\ln A$ distribution for the Mathewson et al \cite{Mathewson1992}
sample with auto-folding and original Mathewson et al TF scaling, but without using the
$\ln A$ algorithm; 
Vertical dotted lines indicate peak centres of Persic \& Salucci solution.
Bin width = 0.055}
\label{FIG17}
\end{figure}
\end{center}

The typical consequences of {\it not} using this algorithm, and computing $\ln A$ on
the whole ORC, are shown in figure \ref{FIG17} which gives the $\ln A$ frequency diagram arising 
from
the Mathewson et al \cite{Mathewson1992} sample after it has been folded using our autofolder
technique but {\it without} using the computational method of \S\ref{TheAlgorithm}.
Again, we see that the discrete states phenomonology is completely obscured.
\subsection{Effect of folding refinements on zero-point determinations for 
scaled ORCs}
\begin{center}
\begin{figure}
\noindent
%\begin{minipage}[h]{\linewidth}
\resizebox{\hsize}{!}{\includegraphics{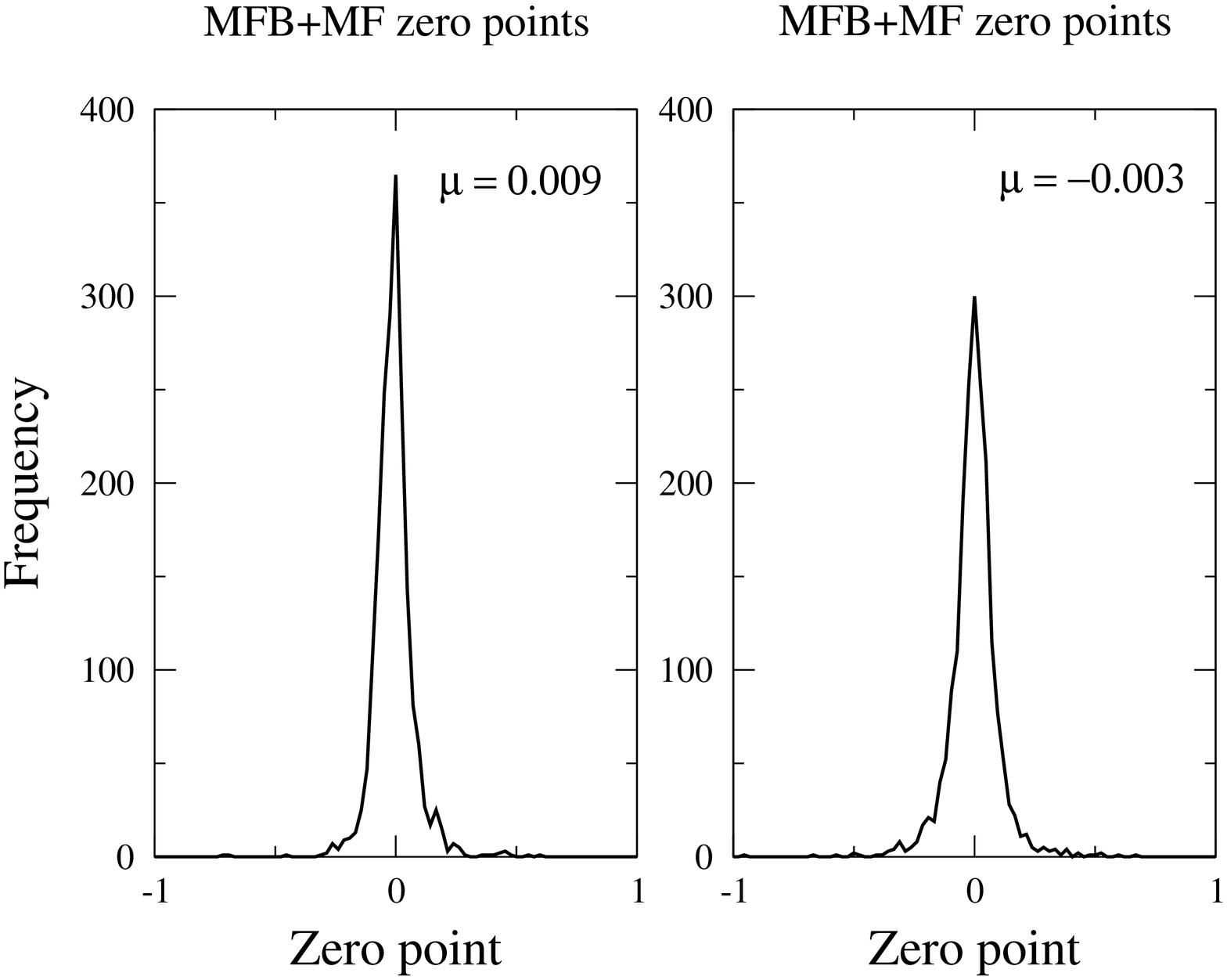}}
%\end{minipage} 
\caption{Left panel: zero-point distribution arising from mfb+mf
combined sample of scaled ORCs with auto-folder and $\ln A$ algorithm.
Right panel: similar to left panel, but with Mathewson's original folding
solution and no $\ln A$ algorithm.
Bin width = 0.055}
\label{FIG18}
\end{figure}
\end{center}

It is of interest to consider the zero-point distribution arising from 
regressing the scaled velocities, $\ln (V/V_0)$ on the scaled radial 
coordinates, $\ln (R/R_0)$, when the $(R_0,\,V_0)$ model is obtained from
ORCs folded using simply $V_{sys}$ determinations and without applying the
$\ln A$ algorithm of \S\ref{TheAlgorithm}.
To illustrate the point, the right panel of figure \ref{FIG18} show the 
distribution arising from
the combined Mathewson et al \cite{Mathewson1992} and Mathewson \& Ford
\cite{Mathewson1996} samples using their original folding solutions, and without
using the $\ln A$ algorithm.
The left panel, included for comparison, arises from the same combined sample,
but with our various folding refinements included.

It is obvious from the figure that the effect of using the original folding
solutions and omitting our refinements is simply to add unbiased noise to the
zero-point solutions.
However, it is interesting to note that, even when the original folding solutions are used
and our various refinements are omitted, the zero-point frequency distribution still provides
strong support for the power-law description of ORCs.
\section{General dynamical implications}
\label{GenDynImp}
The present analysis has allowed us to
established to a very high degree of statistical certainty that
the parameter $\ln A$ appears constrained to take on discrete values 
$k_1, k_2, ...$.
Combining this with, say, (\ref{eqn7A}) which showed
$\ln A \approx F(M,\,S,\,\alpha)$, then we must have 
$F(M,\,S,\,\alpha) = k_1, k_2, ...$.
Thus it appears that spiral galaxies are constrained to exist on one of a set
of discrete class planes in the three-dimensional $(M,S,\alpha)$ space.
This then gives rise to one of two broad possibilities: at some stage 
in its evolution a spiral galaxy somehow moves onto one of these discrete class
planes and then:
\begin{itemize}
\item
is either necessarily constrained to remain on this plane over
the whole of its evolution;
\item 
or has the possibility of transiting to other planes in very short periods 
of time, so that the planes themselves represent an evolutionary sequence.
\end{itemize}
In either case, the primary difficulty is identifying mechanisms which
might generate such discrete sets of possible dynamical class planes and, in the
following, we make three conjectures.
\subsection{First conjecture}
In the wider world on non-linear dynamical systems, it is not difficult to find (or to construct)
systems for which consistent solutions can only be found when certain {\it algebraic} conditions
are satisfied - not eigenvalue problems, but analogous to them.
If, in such a system, the algebraic condition had the form of a quartic defined in the system's
parameter space then, immediately, one would have a situation in which the solutions fell into 
four discrete dynamical classes.
Our first conjecture is, therefore, that disc dynamics are governed by just such a non-linear
system so that the discrete dynamical class phenomonology is merely a manifestation of an unknown
algebraic consistency condition.
\subsection{Second conjecture}
Within the context of inflationary cosmologies, scalar fields play a critical
role in the early universe.
Since it is not difficult to induce oscillatory behaviour in these scalar 
fields, it becomes natural to consider the possibility that the four dynamical
classes of our analysis are the frozen imprint of four distinct galaxy-formation
epochs in the early universe, each associated with a distinct
oscillation of the scalar field.
If this is the case, then the phenomonology provides a very strong constraint on
these early galaxy formation processes.
\subsection{Third conjecture}
Until recently, the most popular candidate for constituting the dark matter 
that is required
to produce the generality of observed rotation curve shapes (especially for
the dimmer galaxies, and beyond optical limits in general) has been the 
neutralino. However, this conjecture is now in serious difficulties since simulations
show that these massive particles would tend to accumulate in large agglomerations
which would cause serious disruption of galactic structure which is not observed.
In response to this state of affairs, Arbey et al \cite{Arbey} have proposed
the existence of a massive non-self interacting charged scalar field as an alternative
source of {\lq dark-matter'} action.
They provide a fairly comprehensive analysis to show how their conjecture provides
extremely good fits to the six brightest of Persic et al's \cite{Persic1996} 
eleven classes
of universal rotation curves.

More particularly, from our point of view, the conjectured dark halo of 
Arbey et al \cite{Arbey}
is based on a
primitive {\lq boson star'} model, and this model is constrained so that it gives
rise to a sequence of discrete energy eigenstates, each of which is associated with
distinct measures of rotation for the {\lq boson star'}.
Whilst it is true that these discrete rotation states do not correspond in any
direct way to the discrete dynamical classes discussed in this paper, it must
be considered remarkable that such a prediction arises at all in a paper which
was primarily directed towards attempting to model the universal ORCs of
Persic et al \cite{Persic1996}.
It seems, therefore, that the phenomonology discussed here may potentially
be understood in terms of the quitessential galactic haloe.
\section{Summary and Conclusions}
\label{Conclusions.sec}
We have analysed four separate large ORC samples to show that the discrete dynamical classes
hypothesis for galaxy discs is supported by the data at the level of virtual certainty.
The immediate significance of the phenomonology is that any given spiral galaxy 
appears to be constrained to evolve over one of a discretely defined set
of dynamical class planes, existing in a three-dimensional $(M,S,\alpha)$ space
where $M$ is absolute magnitude, $S$ is surface brightness and $\alpha$
is a parameter computed for each galaxy from its rotation curve.

We have shown how the broader analysis implies the existence of a second generation of
Tully-Fisher methods which augment the classical Tully-Fisher relationship with a second,
similar, relationship which defines where on an ORC the linewidth is to be measured.
An algorithm for the absolute calibration of this extended Tully-Fisher method was then
proposed.

We have conjectured three possible mechanism for this phenomonology, one 
based on the idea that the phenomonology is the manifestation of an algebraic consistency
condition in some (unkown) non-linear dynamical system theory which describes disc dynamics,
a second based on
the notion of a sequence of distinct galaxy-formation epochs in an oscillating
early universe, and a third based on the dynamical effects of
quintessential halos acting as the source of dark matter around spirals.

Whatever the truth of the matter, it seems certain that the existence of the
distinct dynamical classes
poses very difficult questions for the standard galaxy formation
theories, and will have a a potentially profound affect on our developing 
understanding of galactic dynamics and evolution in particular, 
and the cosmos in general.
\hfill \break
\leftline{\bf Acknowledgements}
I am grateful to M. Persic and P. Salucci, of SISSA Italy, D. Mathewson (now retired) and 
V. Ford of Mt Stromlo
Observatory ANU, D. Dale of the California Institute of Technology,
R. Giovanelli \& M. Haynes of Cornell University and S. Courteau of the 
Herzberg Institute of Astrophysics
Canada, for making available their data, and for patiently answering all
queries, and to Bill Napier of Armagh Observatory, UK, for many fruitful 
discussions during the course of this work. 
\appendix
\section{The Effects of Tully-Fisher Scatter on $\ln A$ Profiles}
\label{Scatter.sec}
All the distance scaling in this analysis is performed using Tully-Fisher
methods, which possess well-understood inherent sources of error. 
Consequently, we need to understand the extent to which these errors can affect
the phenomenon which is the subject of the present analysis.

Mathewson et al \cite{Mathewson1992} report a magnitude scatter of about 0.32 
for their sample (our best)
which compares favourably with that of 0.35 reported by Courteau for his
sample.
These correspond to a scatter of less than 20\% for distance measurements and,
in the following, we analyse the effects of such uncertainties on our
proposed analysis to demonstrate that they cannot wash out any potential
peak structures of the type seen in figure \ref{fig1A}.

Suppose that each galaxy in the sample has had its distance exactly determined,
and that $R$ denotes the corresponding exact radial scale.
Then $V = AR^\alpha$ implies
\begin{displaymath}
\ln V = \ln A + \alpha \ln R.
\end{displaymath} 
The existence of uncertainties in the Tully-Fisher
distance scale with a typical scatter
of 20\% can be accounted for by the replacement $R \rightarrow kR$
where $0.8 \leq k \leq 1.2$, so that 
\begin{displaymath}
\ln V = \ln A + \alpha \ln k + \alpha \ln R \equiv \ln A' + \alpha \ln R,
\end{displaymath} 
where $\ln A' \equiv \ln A + \alpha \ln k$.
We immediately see that uncertainties in the distance scale affect the
zero point in the $(\ln R, \ln V)$ relationship, but leave the gradient
$\alpha$ {\it unaffected}.

Since $0.8 \leq k \leq 1.2$ then, as an approximation, $-0.2 < \ln k < 0.2$
so that $\ln A' \equiv \ln A \pm 0.2 \alpha$.
The peak structures of figure \ref{fig1A} lie in the approximate range 
$3.9 < \ln A < 5.1$ (that is, $1.7 < \log A < 2.2$) and reference to Roscoe
\cite{RoscoeB} (figure 8) shows that this corresponds to the approximate range
$0.18 < \alpha < 0.55$, where low $\alpha$ corresponds to the brightest
galaxies and vice versa.
Therefore, for the brightest galaxies, we have $\ln A' \equiv \ln A \pm 0.04$
whilst, for the dimmest galaxies, we have $\ln A' \equiv \ln A \pm 0.1$.
That is, uncertainties of 20\% in the Tully-Fisher 
distance scale create uncertainties
in $\ln A$ of $\pm 0.04$ at the bright end, and of $\pm 0.1$ at the dim end.

It follows that,
since the mean separation of the $\ln A$ peaks in figure \ref{fig1A} is about 
0.4, these uncertainties in the Tully-Fisher distance scale are
incapable of washing out the discrete peak structure observed in the $\ln A$
distribution.
This analysis provides the confidence required to analysise further samples
on the same basis.
\section{The elimination of artifact as a mechanism}
\label{Artifact.sec}
It is necessary to be as certain as is possible that the phenomonology being
claimed is not created
as an artifact of any particular procedure or data sample. 
There are three independent routes by which such an artifact (no matter how 
remote the possibility) could infiltrate the process, and these can be listed 
as:
\begin{itemize}
\item the original process of measuring ORCs;
\item the method of linewidth estimation;
\item the folding process.
\end{itemize}
\subsection{The original measuring process}
The possibility that an artifact can enter via the ORC measurement process
is minimized by the fact that we analyse four different samples originating
with three distinct groups of astronomers using three different telescopes in 
different hemispheres.
A detailed discussion of the samples is given in \S\ref{Sec2}.
\subsection{Linewidth determinations}
\label{LWDs}
The possibility that an artifact can enter via the linewidth estimation process
is minimized by the fact that five different methods of linewidth determination are employed 
over our four samples: 
\begin{itemize}
\item
The Mathewson et al \cite{Mathewson1992} sample has linewidths determined using
a non-algorithmic and intuitive {\lq eye-ball'} method (private communication);
\item the Mathewson \& Ford \cite{Mathewson1996} sample
is considered using (a) Mathewson's own {\lq eye-ball'} linewidth determinations and
(b) algortihmically determined linewidths based on $V_{opt}$, the velocity at the optical
radius as defined by Persic \& Salucci \cite{Persic1995};
\item the Dale et al \cite{Dale1997} sample has linewidths determined by an elaborate process
based on the $V_{90\%}-V_{10\%}$ linewidth estimator, which is 
similar to the technique first introduced by Dressler \& Faber \cite{Dressler};
\item the Courteau sample was taken from the study of Courteau \cite{Courteau} which was explicitly 
designed as a study
of objective black-box methods of optical linewidth estimation.
He tests a variety of methods, and we present results using those two which he
judges to be the best, $V_{2.2}$, and the worst, $V_{max}$, respectively.
\end{itemize}
\subsection{The folding process}
Finally, the possibility that an artifact can enter via the folding process is
minimized by the fact that the phenomenon is observed when either of two quite
distinct folding methods is used.
These two methods, described below, share the features introduced by Persic \& Salucci
\cite{Persic1995} that (a) the folding process requires the determination of {\it two} 
parameters, the primary one being 
$V_{sys}$ the systematic velocity (which is usually the only one considered), and the 
secondary one being $\Delta \Phi$ which measures the angular offset between the optical and
the kinematic centres and (b) the pre-folding data filter described in \S\ref{PFDF}.
These features are considered necessary for the purpose of maximising folding accuracy.
\subsubsection{The method of Persic and Salucci}
Persic \& Salucci \cite{Persic1995} were primarily interested in using rotation curves
for studies of the interior dynamics of spiral galaxies and so, by their own
criteria, had a requirement for a large sample of particularly accurately 
folded ORCs.
They took the 965 ORCs of Mathewson et al \cite{Mathewson1992} and used an 
eye-ball method of folding
to produce a sample of 900 good-to-excellent quality 
folded ORCs; as a qualitative measure of the effort expended to produce
this sample, we can note that it took these two authors about a year to
process it (private communication).
Every velocity measurement in the Mathewson et al \cite{Mathewson1992} sample 
came provided with a parameter
(varying on the range $(0,\,1)$) which estimated the relative internal
accuracy associated with the measurement.
Persic \& Salucci \cite{Persic1995} found that the accurate folding of any given ORC required the {\it rejection}
of any individual velocity measurement for which the associated accuracy 
parameter was $\leq\, 0.35$.
In the present context, only the Mathewson et al \cite{Mathewson1992} sample 
has been folded with this method.
\subsubsection{The auto-folder method of Roscoe \cite{RoscoeC}}
\label{Sec.Autofolder}
This method was developed in anticipation of the need accurately to fold
the Mathewson \& Ford \cite{Mathewson1996} sample of 1200+ ORCs on a reasonable time-scale.
The details of this method are described in Roscoe \cite{RoscoeC} but, briefly, it is based on 
the formal minimization of the symmetric components in Fourier representations 
of ORCs with respect to variations in the two folding parameters.

The folding method was developed on that subset of the 
Mathewson et al \cite{Mathewson1992} 
sample used by Persic \& Salucci \cite{Persic1995} and, corresponding to the 
experience of these latter authors, we found that the optimal trade-off 
point between the quality 
of individual velocity measurements, and the volume of good-quality data 
available for the automatic folding method, required the prior rejection of any 
individual velocity measurement which had an associated relative accuracy 
parameter $\leq \, 0.4$.
This is roughly equivalent to the requirement that the absolute
error of any given velocity measurement should be $ \leq \,5\%$.

This folding method has been used here on the samples of 
Mathewson et al \cite{Mathewson1992}, Mathewson \& Ford \cite{Mathewson1996}, 
Dale et al \cite{Dale1997} et seq and Courteau \cite{Courteau}.
\section{Overview of the ORC dynamical partitioning process}
\label{RminRopt}
We discuss two methods of assessing 
the efficiency and effectiveness of the process described in
\S\ref{TheAlgorithm} for the computation of $\ln A$ and establish, with virtual certainty,
the truth of the statements that: 
\begin{itemize}
\item the innermost parts of ORCs exhibit behaviour which is qualitatively
sharply distinguished from that exhibited by the outermost parts of ORCs;
\item the size of such innermost sections can be quantified in terms of a 
radial measure, $R_{min}$ say, which can 
be shown to be extremely powerfully correlated
with the independently defined optical radius, $R_{opt}$ 
(here, as given by Persic \& Salucci \cite{Persic1995} of the disc.
Since $R_{opt}$ carries physical information about the system, then we must
conclude that the algorithmically estimated $R_{min}$ likewise carries physical
information about the system.
Given the quality of the statistics involved, 
these two points are entirely sufficient to establish that $R_{min}$ 
does, in fact, define a real boundary between distinct
dynamical regimes which, in turn, gives a
concrete justification to the technique by which it is estimated.
We suggest that $R_{min}$ could act as a tracer for the gravitational radius
of the core on the basis of the circumstance 
that there appears to be no other possible interpretation.
\end{itemize}
\begin{figure}{h}
\resizebox{\hsize}{!}{\includegraphics{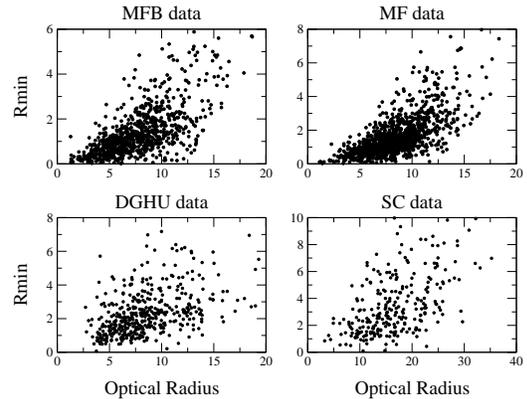}}
\caption{Scatter plot of $R_{opt}:R_{min}$ for MFB, MF, DGHU and SC data with dynamical
partitioning}
\label{fig12A}
\end{figure}
\subsection{The first test of dynamical partitioning}
We argued in Roscoe \cite{RoscoeB} that, if $R_{min}$ really 
was a tracer for the gravitational radius of the core, then
we might expect to find a positive correlation between $R_{min}$ and $R_{opt}$ -
on the grounds that galaxies with large cores might be expected to have large
optical radii etc.

The top-left panel of figure \ref{fig12A} shows the ($R_{opt}:R_{min}$) scatter plot for Mathewson 
et al data \cite{Mathewson1992}, and we note the existence of
an extremely strong positive ($R_{opt}:R_{min}$) correlation.
The top-right panel shows an equally strong effect for 
Mathewson \& Ford \cite{Mathewson1996} data, whilst the bottom-left panel 
shows a
similarly strong effect for the Dale et al \cite{Dale1997} et seq sample,
and and the bottom-right panel shows the same effect for the much
smaller Courteau sample.
The statistics underlying these diagrams are given in the various columns of
table \ref{Table5} and 
these confirm in quantitative terms what is obvious in the diagrams.
We can therefore conclude with certainty that, since $R_{opt}$ is an objectively
defined physical boundary, the $R_{min}$ is similarly an objectively defined
physical boundary. We speculate that, in fact, it is a noisy tracer of the
boundary between core-dominated and disc-dominated dynamics in the disc.
\begin{table*}
\caption{$R_{opt} = b_0 + b_1 \,R_{min} $}
\label{Table5}
\begin{tabular}{lll|ll|ll|ll}
&\multicolumn{2}{c}{MFB} & 
\multicolumn{2}{c}{ MF}&
\multicolumn{2}{c}{ DGHU}&
\multicolumn{2}{c}{ SC} \\
Predictor & Coeff  & t-ratio& Coeff  & t-ratio& Coeff  & t-ratio& Coeff  & t-ratio \\
\hline
$Const.$    & 5.26 & 44&  5.95 & 55& 5.97 & 23& 12.45 & 25 \\
$R_{min}$ & 1.74 & 28&  1.54 & 30&  1.06 & 11& 1.21 & 12  \\
\hline
&\multicolumn{2}{c}{$R^2$ ~=~ 48.9\%} & 
\multicolumn{2}{c}{ $R^2$ ~=~ 45.7\%}&
\multicolumn{2}{c}{ $R^2$ ~=~ 22.2\%}&
\multicolumn{2}{c}{ $R^2$~=~ 35.5\%} \\
\hline
\end{tabular}
\end{table*}
The foregoing considerations lead to the following conclusions:
\begin{itemize}
\item 
The application of the dynamical partitioning process produces a very
powerful, although noisy, $R_{min}:R_{opt}$ correlation
confirming that $R_{min}$ (as computed by dynamical partitioning) is a 
strong tracer for $R_{opt}$;
\item The computed value of $R_{min}$ defines a physical transition
boundary between core-dominated dynamics and disc-dominated dynamics.
\end{itemize}
Taking these items together, and noting the absence of any other obvious 
interpretation, we conclude that $R_{min}$ almost certainly represents a 
dynamically derived tracer of the gravitational radius of the core.
\begin{table*}
\caption{Effect of dynamical partitioning technique on power-law fits}
\label{Table4}
\begin{tabular}{lllll}
Dynamical & MFB & MF &DGHU & SC \\
partitioning &Mean rms& Mean rms &Mean rms& Mean rms\\
\hline
Before & $10.8 \times 10^{-2}$ & $12.0 \times 10^{-2}$ & $1.1 \times 10^{-2}$ & 
$ 1.8 \times 10^{-2}$ \\
After & $~2.4 \times 10^{-2}$ & $~3.4 \times 10^{-2}$& $ 0.6 \times 10^{-2}$ &
$1.4 \times 10^{-2}$ \\
\hline
\multicolumn{4}{l}{Statistics calculated on the 866 ORCs of MFB, the 1085 ORCs
of MF}\\
\multicolumn{4}{l}{the 454 ORCs of DGHU and the 283 ORCs of SC.}\\
\hline
\end{tabular}
\end{table*}
\subsection{The second test of dynamical partitioning}
The second definitive formal measure of the statistical efficiency of dynamical partitioning
is given in table \ref{Table4}.
The first row of this table gives the averaged root mean square (rms) error 
calculated from 
fitting power-laws to each of the 866 foldable ORCs of the 
Mathewson et al \cite{Mathewson1992} sample, the 1085 foldable ORCs of the 
Mathewson \& Ford \cite{Mathewson1996} sample, the 454 foldable ORCs of the Dale et al \cite{Dale1997}
et seq sample and the 283 foldable ORCs of the Courteau \cite{Courteau} sample 
{\it before} the dynamical partitioning process.
The second row gives the corresponding averaged rms values {\it after} the 
dynamical partitioning process.
Column 2 refers to Mathewson et al \cite{Mathewson1992} data, column 3 
refers to Mathewson \& 
Ford \cite{Mathewson1996} data, coulumn 4 refers to Dale et al \cite{Dale1997} et seq data
and column 5 refers to Courteau \cite{Courteau} data.
The Mathewson et al \cite{Mathewson1992} data shows an almost 80\% reduction 
in its mean rms, the
Mathewson \& Ford \cite{Mathewson1996} data shows a 72\% reduction in its mean rms,
the Dale et al \cite{Dale1997} et seq data shows a 45\% reduction in its mean rms and the
Courteau \cite{Courteau} data shows a 29\% reduction in its mean rms.

Since the dynamical partitioning process leads to the discarding of only 12\% of the 
Mathewson et al \cite{Mathewson1992} and the Mathewson \& Ford \cite{Mathewson1996} data 
(out of a total of about 37000 individual measurements over the two  samples),
of only
9\% of the Dale et al \cite{Dale1997} et seq data (out of a total of 15000 individual 
measurements)
and of only 10\% of the Courteau \cite{Courteau} data (out of a total of about 17000 individual measurements), 
then we can categorically state that the table
provides conclusive evidence for the statement that the deviation from the 
power-law fit is strongly concentrated on the inner 10\% or so of ORCs.
From this we can conclude that the behaviour of the inner 10\% or so 
of ORCs is qualitatively sharply distinguished from the outer 90\% or so, as we would 
expect on purely dynamical grounds. 
\section{Confidence limits on peak positions}
\label{sec.ConfidenceLims}
\begin{figure}
\noindent
\begin{minipage}[b]{\linewidth}
\resizebox{\hsize}{!}{\includegraphics{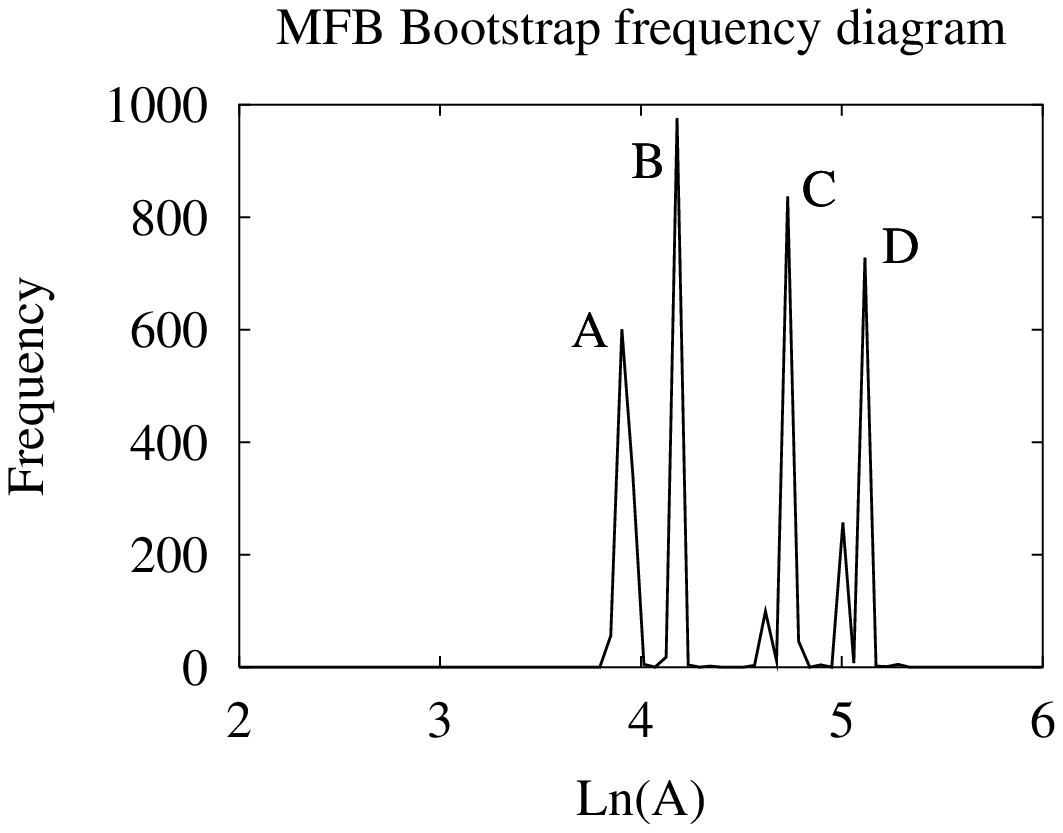}}
\end{minipage} 
\caption{Frequency diagram for the positions of the largest signals near the
original $A, B, C, D$ peaks, bootstrapped from the MFB sample over 1000 trials}
\label{fig13}
\end{figure}

We use bootstrap techniques to show that the uncertainties in the positions of 
the $A, \,B,\, C,\, D$ peaks of figure \ref{fig1}, are small.
The process adopted is described as follows:
\begin{itemize}
\item Partition the range of $\ln A$ into the sub-ranges
$(2.200,4.070)$ containing peak $A$, $(4.070,4.455)$ containing peak $B$,
$(4.455,4.950)$ containing peak $C$ and $(4.950,6.000)$ containing peak $D$;
\item Use the actual $\ln A$ data-set of $N$ distinct $\ln A$ values to generate
1000 bootstrapped simulated data-sets, each consisting of $N$ values generated 
by random selection from the real data-set with replacement;
\item Form the frequency diagram for each of the 1000 simulated data-sets, 
and record the position of the 
{\it largest} signal only in each of the four sub-ranges defined above;
\item Form the frequency diagram for this {\lq largest signal'} data set -
this is plotted in figure \ref{fig13}.
\end{itemize}  
We now discuss the peaks in figure \ref{fig13} in turn: 
\hfill \break
{\bf Peak $A$: $\ln A = 3.91$}  \hfill \break
This peak coincides exactly with peak $A$ of figure
\ref{fig1}.
Since $\ln A$ correlates in a strong positive sense
with absolute luminosity (Roscoe \cite{RoscoeB}, then this peak corresponds to the
dimmest end of the sample, so that greater measurement uncertainties probably
account for the relative broadness of this peak; even so, this peak is still
very tightly defined and the 90\% confidence interval this peak is
(3.87,3.98).
\hfill \break
{\bf Peak $B$: $\ln A = 4.18$} \hfill \break
This peak coincides exactly with peak $B$ of figure
\ref{fig1} and is the most straightforward case, being very
tightly defined with no subsidiary peaklets; in fact, more than 97\% of the
sampled $B$ peaks fall in a single bin, so that we can assert that an extremely
conservative 90\% confidence interval for this peak is given by the bin
boundaries as (4.15,4.21). 
\hfill \break
{\bf Peak $C$: $\ln A = 4.73$} \hfill \break
This peak exhibits a bi-modal structure with the major mode coinciding exactly
with the peak $C$ of figure \ref{fig1}; the minor peak contains about 10\% of
the sampled $C$ peaks, with the remainder being in the major peak.
The boundaries of the major peak, given by (4.70,4.76), therefore give an 
estimate of a 90\% confidence interval for the $C$ peak.
\hfill \break
{\bf Peak $D$: $\ln A = 5.12$} \hfill \break
Again there is a bi-model structure, but this time the minor mode is very much
more prominent than it is for the $C$ peak.
We find that the minor peak contains approximately 20\% of the sampled $D$
peaks, with the remainder being in the major peak.
To simplify the complexities presented by this bi-modal structure, we simply 
choose the boundaries of the major peak as estimates of the required 90\%
confidence interval for the $D$ peak.
%
% SC ORCs have on average three times as many measurements per ORC 
% as do Mathewson, Ford \& Buchhorn (1992) ORCs (95:30).
% Setting error = 5 per cent == to using best 55 per cent of velocity data!

\end{document}